\documentclass[aps,prc,showpacs,preprintnumbers,superscriptaddress,nofootinbib,floatfix,10pt]{revtex4-1}
\usepackage{amsfonts,amssymb,stmaryrd,latexsym,amsmath}
\usepackage{subfigure,graphicx}
\usepackage{times}
\usepackage{slashed}
\usepackage[colorlinks]{hyperref}

\usepackage{graphicx}
\usepackage{amsmath, amssymb, bm}
\usepackage{dcolumn}
\usepackage{bm}
\usepackage{comment}
\usepackage{tabularx}
\usepackage{multirow}

\begin{document}

\title{Neutron-proton scattering with lattice chiral effective field theory \\ at next-to-next-to-next-to-leading order}

\author{Ning~Li}
\email{lini@frib.msu.edu} 
\affiliation{Facility for Rare Isotope Beams and Department of Physics and
Astronomy, Michigan State
University, MI 48824, USA}

\author{Serdar~Elhatisari}
\affiliation{Helmholtz-Institut f\"ur Strahlen- und
             Kernphysik and Bethe Center for Theoretical Physics,
             Universit\"at Bonn,  D-53115 Bonn, Germany}
\affiliation{Department of Physics, Karamanoglu Mehmetbey University,
Karaman 70100, Turkey}

\author{Evgeny~Epelbaum}
\affiliation{Institut~f\"{u}r~Theoretische~Physik~II,~Ruhr-Universit\"{a}t~Bochum,
D-44870~Bochum,~Germany}

\author{Dean~Lee}
\affiliation{Facility for Rare Isotope Beams and Department of Physics and Astronomy, Michigan State
University, MI 48824, USA}
\affiliation{Department~of~Physics, North~Carolina~State~University, Raleigh,
NC~27695, USA}
                     
\author{Bing-Nan~Lu}
\affiliation{Facility for Rare Isotope Beams and Department of Physics and
Astronomy, Michigan State
University, MI 48824, USA}

\author{Ulf-G.~Mei{\ss}ner}  
\affiliation{Helmholtz-Institut f\"ur Strahlen- und
             Kernphysik and Bethe Center for Theoretical Physics,
             Universit\"at Bonn,  D-53115 Bonn, Germany}  
\affiliation{Institute~for~Advanced~Simulation, Institut~f{\"u}r~Kernphysik,
and
J{\"u}lich~Center~for~Hadron~Physics,~Forschungszentrum~J{\"u}lich,
D-52425~J{\"u}lich, Germany}     
\affiliation{JARA~-~High~Performance~Computing, Forschungszentrum~J{\"u}lich,
D-52425 J{\"u}lich,~Germany}

\begin{abstract}  
 We present a new lattice formulation of chiral effective field theory interactions with a simpler decomposition into spin channels.  
 With these interactions the process of fitting to the empirical scattering phase shifts is simplified, and the resulting lattice phase 
 shifts are more accurate than in previous studies.  We present results for the neutron-proton system up to next-to-next-to-next-to-leading 
 order for lattice spacings of $1.97$, $1.64$, $1.32$, and $0.99~{\rm fm}$.  Our results provide a 
 pathway to \textit{ab initio} lattice calculations of nuclear structure, reactions, and thermodynamics with accurate and systematic 
 control over the chiral nucleon-nucleon force.                
\end{abstract}


\maketitle

\section{Introduction}
 Chiral effective field theory (EFT) organizes the interactions of nucleons in powers of momenta and factors of the pion mass near 
 the chiral limit where the light quarks are massless. We label terms that carry a total of $n$ powers of nucleon momenta or 
 factors of the pion masses as order $Q^n$.  The most important interactions at low energy are at order $Q^0$, or leading order (LO).    
 Next-to-leading order (NLO) interactions correspond to order $Q^2$, next-to-next-to-leading order (N$^2$LO) terms ato $Q^3$, and 
 next-to-next-to-next-to-leading order (N$^3$LO) to  $Q^4$. See  Ref.~\cite{Epelbaum:2008ga} for a review of chiral EFT.  Nuclear lattice 
 simulations using chiral EFT have been
used in recent years to describe the structure and scattering of atomic nuclei
\cite{Epelbaum:2011md,Epelbaum:2012qn,Epelbaum:2013paa,Elhatisari:2015iga,Elhatisari:2016owd,
Elhatisari:2017eno}. However, the treatment of nuclear forces at higher orders in the chiral EFT expansion is more difficult 
on the lattice due to the breaking of rotational invariance produced by  nonzero lattice spacing \cite{Alarcon:2017zcv,Klein:2018lqz}. 
Fitting the unknown coefficients of the short-range lattice interactions to empirical phase shifts can introduce significant uncertainties. 

In this paper we solve these problems by introducing a new set of short-range chiral EFT interactions on the lattice with a simpler 
decomposition into spin channels.  The angular dependence of the relative separation between the two nucleons is prescribed by 
spherical harmonics, and the dependence on the nucleon spins is given by the spin-orbit Clebsch-Gordan coefficients. The full 
details of this process are presented in this paper. We start with some definitions of the lattice operator notations used. Next 
we discuss the lattice Hamiltonian used in our lattice transfer matrix formalism.  The short-range interactions are presented 
first, and we then proceed to the long-range interactions.  We then compare our neutron-proton scattering results at lattice spacings of $1.97$, 
$1.64$, $1.32$, and $0.99~{\rm fm,}$ with the empirical phase shifts.  After this we compute some observable 
properties of the deuteron, discuss theoretical uncertainties, and present a summary and outlook. Certain interactions such as the Coulomb interaction and some  
isospin-breaking interactions are not directly relevant to the neutron-proton analysis that we consider here. However we include these 
interactions in this work for completeness and future reference.

\section{Lattice operator definitions} \label{definition}

Let us define $a_{i,j}({\bf n})$ and $a^{\dagger}_{i,j}({\bf n})$, the lattice
annihilation and creation operators on lattice site ${\bf n}$ with spin $i=0,1$
(up, down) and isospin $j=0,1$ (proton, neutron). The operators $a^{s_{\rm NL}}_{i,j}({\bf n})$ and $a^{s_{\rm NL}\dagger}_{i,j}({\bf n})$ are defined
via nonlocal smearing with the real parameter $s_{\rm NL}$,
\begin{equation}
 a^{s_{\rm NL}}_{i,j}({\bf n})=a_{i,j}({\bf n})+s_{\rm NL}\sum_{|{\bf n'}|=1}a_{i,j}({\bf
n}+{\bf n'}).
\end{equation}
\begin{equation}
 a^{s_{\rm NL}\dagger}_{i,j}({\bf n})=a^{\dagger}_{i,j}({\bf n})+s_{\rm NL}\sum_{|{\bf
n'}|=1}a^{\dagger}_{i,j}({\bf
n}+{\bf n'}).
\end{equation}

Next we define the pair annihilation operators $[a({\bf n})a({\bf n'})]^{s_{\rm
NL}}_{S,S_z,I,I_z}$, where
\begin{equation}
 [a({\bf n})a({\bf n'})]^{s_{\rm
NL}}_{S,S_z,I,I_z}=\sum_{i,j,i',j'} a^{s_{\rm NL}}_{i,j}({\bf n})M_{ii'}(S,S_z)M_{jj'}(I,I_z)a^{s_{\rm
NL}}_{i',j'}({\bf n'})
\label{spin-isospin}
\end{equation}
with
\begin{equation}
M_{ii'}(0,0)=\frac{1}{\sqrt{2}}[\delta_{i,0}\delta_{i',1}-\delta_{i,1}\delta_{i',0}],
\end{equation}
\begin{equation}
M_{ii'}(1,1)=\delta_{i,0}\delta_{i',0},
\end{equation}
\begin{equation}
M_{ii'}(1,0)=\frac{1}{\sqrt{2}}[\delta_{i,0}\delta_{i',1}+\delta_{i,1}\delta_{i',0}],
\end{equation}
\begin{equation}
M_{ii'}(1,-1)=\delta_{i,1}\delta_{i',1}.
\end{equation}
We define the lattice finite difference operation $\nabla_l$ on a general
lattice function $f({\bf n})$ as
\begin{equation}
\nabla_lf({\bf n})=\frac{1}{2}f({\bf n}+{\bf \hat{l}})-\frac{1}{2}f({\bf
n}-{\bf \hat{l}}),
\end{equation}
where ${\bf \hat{l}}$ is the spatial lattice unit vector in the $l$ direction. 
It is also convenient to define the lattice finite difference operation $\nabla_{1/2,l}$
defined on points halfway between lattice sites. 
\begin{equation}
\nabla_{1/2,l}f({\bf n})=f({\bf n}+\frac{1}{2}{\bf \hat{l}})-f({\bf
n}-\frac{1}{2}{\bf \hat{l}}).
\end{equation} 
We use this operation only to define the Laplacian operator,
\begin{equation}
{\bf \nabla}^{2}_{1/2} = \sum_l{\bf \nabla}^{2}_{1/2,l}.
\end{equation}
Let us define the solid harmonics
\begin{equation}
R_{L,L_z}({\bf r}) = \sqrt{\frac{4\pi}{2L+1}}r^L Y_{L,L_z}(\theta,\phi),
\end{equation}
and their complex conjugates
\begin{equation}
R^*_{L,L_z}({\bf r}) = \sqrt{\frac{4\pi}{2L+1}}r^L Y^*_{L,L_z}(\theta,\phi).
\end{equation}
Using the pair annihilation operators, lattice finite differences,
and solid harmonics, we define
the operator
\begin{equation}
P^{2M,s_{\rm NL}}_{S,S_z,L,L_z,I,I_z}({\bf n}) =[a({\bf n}){\bf \nabla}^{2M}_{1/2}R^*_{
L,L_{z}}({\bf \nabla})a({\bf n})]^{s_{\rm
NL}}_{S,S_z,I,I_z}, 
\end{equation}
where ${\bf \nabla}_{1/2}^{2M}$ and ${\bf \nabla}$ act on the second
annihilation operator.  More explicitly stated, this means that we act on
the ${\bf n'}$ in Eq.~(\ref{spin-isospin}) and then set ${\bf n'}$ to equal
${\bf n}$.
The even integer $2M$ gives us higher powers of the finite differences. Writing the Clebsch-Gordan coefficients as $\langle S S_z, L L_z \vert J J_z\rangle$,
we define 
\begin{equation}
O^{2M,s_{NL}}_{S,L,J,J_z,I,I_z}({\bf n})= 
\sum_{S_z,L_z}\langle S S_z, L L_z \vert J J_z\rangle P^{2M,s_{NL}}_{S,S_z,L,L_z,I,I_z}({\bf
n}).
\end{equation}

We also define point-like density operators that depend on spin and isospin.  For spin indices $S=1,2,3,$ and isospin indices $I=1,2,3$, we define\begin{align}
\rho({\bf n})&= \sum_{i,j}a^\dagger_{i,j}({\bf n}) a_{i,j}({\bf n}),
\\
\rho_{S}({\bf n})&=\sum_{i,i',j}a^\dagger_{i,j}({\bf n})[\sigma_S]_{ii'} a_{i',j}({\bf
n}),
\\
\rho_{I}({\bf n})&=\sum_{i,j,j'}a^\dagger_{i,j}({\bf n})[\tau_I]_{jj'} a_{i,j'}({\bf
n}),
\\
\rho_{S,I}({\bf n})&=\sum_{i,i',j,j'}a^\dagger_{i,j}({\bf n})[\sigma_S]_{ii'} \otimes
[\tau_I]_{jj'} a_{i
',j'}({\bf n}),
\end{align}
where $\sigma_S$ are Pauli matrices in spin space  and $\tau_I$ are Pauli matrices in isospin space.
\section{Lattice Hamiltonian and transfer matrix formalism} \label{Hamiltonian}
Let $a$ be the spatial lattice spacing and $a_t$ be the temporal lattice spacing.  We work in lattice units (l.u.) where all quantities are 
multiplied by the powers of the spatial spacing to form a dimensionless combination. The normal-ordered transfer matrix is
\begin{equation}
M = : \exp[-H\alpha_t] :, 
\end{equation}
where the $::$ symbols denote normal ordering where the annihilation operators are on the right and creation operators are on the left. 
$\alpha_t = a_t/a$ is the ratio between the temporal lattice spacing and the spacial lattice spacing. 
For the temporal lattice spacing, we take $a_t = 1.32~{\rm fm}$ for $a = 1.97~{\rm fm}$. We rescale 
$a_t$ for other lattice spacings so that $a^2/a_t$ is fixed.
We partition the lattice Hamiltonian $H$ into a free Hamiltonian, short-range interactions, and long-range interactions,
  
\begin{equation}
H_{\rm }^{\rm } =H_{\rm free} +V^{\rm short}_{\rm 2N} +\ V^{\rm long}_{\rm
2N}. 
\end{equation}For the free Hamiltonian we use an $O(a^4)$-improved action of the form  \cite{Lee:2008fa},
\begin{align}
H_{\rm free}= &\frac{49}{12m}\sum_{\bf n} a^\dagger({\bf n}) a({\bf n})-\frac{3}{4m}\sum_{{\bf
n},i}
\sum_{\langle{\bf n'}\,{\bf n}\rangle_i} a^\dagger({\bf n'}) a({\bf n}) \nonumber
\\
&+\frac{3}{40m}\sum_{{\bf
n},i}\sum_{\langle\langle{\bf n'}\,{\bf n}\rangle\rangle_i} a^\dagger({\bf
n'})
a({\bf n})-\frac{1}{180m}\sum_{{\bf
n},i}
\sum_{\langle\langle\langle{\bf n'}\,{\bf n}\rangle\rangle\rangle_i} a^\dagger({\bf
n'}) a({\bf n}).
\end{align}
\section{short-range interactions} \label{short:range}
\subsection{Order $Q^0$}
At order $Q^0$ we have two short-range interaction operators, namely, the $S$-wave spin
singlet which we call $V_{0,^1S_0}({\bf n})$, 
\begin{equation}
\sum_{I_z=-1,0,1}\left[ O^{0,s_{NL}}_{0,0,0,0,1,I_z}({\bf n})\right]^{\dagger}O^{0,s_{NL}}_{0,0,0,0,1,I_z}({\bf
n})  
\end{equation}
and the $S$-wave spin triplet $V_{0,^3S_1}({\bf n})$,
\begin{equation}
\sum_{J_z=-1,0,1}\left[ O^{0,s_{NL}}_{1,0,1,J_z,0,0}({\bf n})\right]^{\dagger}O^{0,s_{NL}}_{1,0,1,J_z,0,0}({\bf
n}). 
\end{equation}
We note that since we work with interactions that act with a specified parity
and specified total intrinsic spin $S$, when we act on two-nucleon states
with total momentum equal to zero, the
total isospin $I$ is completely constrained by the requirement
of overall antisymmetry of the two nucleons. However we still specify
the correct total isospin $I$ explicitly in order to remove lattice artifacts
that might otherwise appear in cases when the total momentum is not zero.

Wigner's SU(4) symmetry \cite{Wigner:1936dx}
is an approximate symmetry of the low-energy nucleon-nucleon interactions
where the nucleonic spin and isospin degrees of freedom can be rotated as
four components of an SU(4) multiplet.  As in previous work \cite{Elhatisari:2017eno},
we treat the SU(4) part of the short-range interactions separately.
This choice allows us to control the strength of the local part of the
SU(4) interaction, which has been shown to be important for  the binding of nucleons
in nuclei \cite{Elhatisari:2016owd}. So at leading order we also include an SU(4)-invariant short-range operator $V_0$
with the form
\begin{equation}
V_0 = \frac{C_0}{2}: \! \! \! \sum_{{\bf n'},{\bf n},{\bf n''}}\sum_{i',j'}
a^{s_{\rm NL}\dagger}_{i',j'}({\bf
n'})a^{s_{\rm NL}}_{i',j'}({\bf n'})
f_{s_{\rm L}}({\bf n'} - {\bf n})f_{s_{\rm L}}({\bf n} - {\bf n''})
\sum_{i'',j''} a^{s_{\rm NL}\dagger}_{i'',j''}({\bf n''})a^{s_{\rm NL}}_{i'',j''}({\bf
n''}):,
\end{equation}
where $f_{s_{\rm L}}$ is defined as
\begin{align}
f_{s_{\rm L}}({\bf n})& = 1 \; {\rm for} \; |{\bf n}| = 0, \nonumber \\
& = s_L \; {\rm for} \; |{\bf n}| = 1, \nonumber \\
& = 0 \; {\rm otherwise}.
\end{align} 
We repeat again that, in terms of counting powers of momentum, this SU(4) interaction is equivalent to the SU(4)-invariant interaction we get by adding together  $V_{0,^1S_0}({\bf
n})$ and $V_{0,^3S_1}({\bf
n})$.  However, the separate treatment of this interaction allows us to control the strength of the local part of the
SU(4) interaction in systems with more than a few nucleons.  For the purposes of fitting operator coefficients, 
we keep the coefficient $C_0$ fixed and tune the coefficients of $V_{0,^1S_0}({\bf n})$ and $V_{0,^3S_1}({\bf n})$ 
as needed to reproduce the scattering phase shifts and mixing angles.  In this work we take the smearing parameter 
$s_{\rm NL}$ for the SU(4) interaction to be the same as that used in the other short-range interactions.
For $a = 1.97~{\rm fm}$ we take $C_0 = -0.175~{\rm l.u.}$, $s_L = 0.070~{\rm l.u}$, and $s_{\rm NL} = 0.080~{\rm l.u.}$.
For $a = 1.64~{\rm fm}$ we use $C_0 = -0.100~{\rm l.u.}$, $s_L = 0.109~{\rm l.u}$, and $s_{\rm NL} = 0.122~{\rm l.u.}$.
For $a = 1.32~{\rm fm}$, we use $C_0 = -0.045~{\rm l.u.}$, $s_L = 0.170~{\rm l.u}$, and $s_{\rm NL} = 0.186~{\rm l.u.}$.
For $a = 0.99~{\rm fm}$, we use $C_0 = -0.015~{\rm l.u.}$, $s_L = 0.265~{\rm l.u}$, and $s_{\rm NL} = 0.283~{\rm l.u.}$.
In future work, however, we may consider different smearing parameters for the two cases in order to accelerate the 
convergence of the effective field theory expansion in many-body systems.

\subsection{Order $Q^2$}
At order $Q^2$ we have the lowest radial excitations of the $S$-wave spin
singlet which we call $V_{2,^1S_0}({\bf n})$,
\begin{equation}
\sum_{I_z=-1,0,1}\left[ O^{2,s_{NL}}_{0,0,0,0,1,I_z}({\bf n})\right]^{\dagger}O^{0,s_{NL}}_{0,0,0,0,1,I_z}({\bf
n}) + \sum_{I_z=-1,0,1}\left[ O^{0,s_{NL}}_{0,0,0,0,1,I_z}({\bf n})\right]^{\dagger}O^{2,s_{NL}}_{0,0,0,0,1,I_z}({\bf
n}), 
\end{equation}
and the $S$-wave spin triplet
$V_{2,^3S_1}({\bf n}),$\begin{equation}
\sum_{J_z=-1,0,1}\left[ O^{2,s_{NL}}_{1,0,1,J_z,0,0}({\bf n})\right]^{\dagger}O^{0,s_{NL}}_{1,0,1,J_z,0,0}({\bf
n})+\sum_{J_z=-1,0,1}\left[ O^{0,s_{NL}}_{1,0,1,J_z,0,0}({\bf n})\right]^{\dagger}O^{2,s_{NL}}_{1,0,1,J_z,0,0}({\bf
n}). 
\end{equation}
At order $Q^2$ there is the $^1P_1$ interaction $V_{2,^1P_1}({\bf n})$, 
\begin{equation}
\sum_{J_z=-1,0,1}\left[ O^{0,s_{NL}}_{0,1,1,J_z,0,0}({\bf n})\right]^{\dagger}O^{0,s_{NL}}_{0,1,1,J_z,0,0}({\bf
n}), 
\end{equation}
the $^3P_0$ interaction $V_{2,^3P_0}({\bf n})$,
\begin{equation}
\sum_{I_z=-1,0,1}\left[ O^{0,s_{NL}}_{1,1,0,0,1,I_z}({\bf n})\right]^{\dagger}O^{0,s_{NL}}_{1,1,0,0,1,I_z}({\bf
n}), 
\end{equation}
the $^3P_1$ interaction $V_{2,^3P_1}({\bf n})$,
\begin{equation}
\sum_{I_z=-1,0,1}\sum_{J_z=-1,0,1}\left[ O^{0,s_{NL}}_{1,1,1,J_z,1,I_z}({\bf
n})\right]^{\dagger}O^{0,s_{NL}}_{1,1,1,J_z,1,I_z}({\bf
n}), 
\end{equation}
and the $^3P_2$ interaction $V_{2,^3P_2}({\bf n})$,
\begin{equation}
\sum_{I_z=-1,0,1}\sum_{J_z=-2,...2}\left[ O^{0,s_{NL}}_{1,1,2,J_z,1,I_z}({\bf
n})\right]^{\dagger}O^{0,s_{NL}}_{1,1,2,J_z,1,I_z}({\bf
n}). 
\end{equation}
At order $Q^2$ we also have the $S-D$ mixing term $V_{2,SD}({\bf n})$,
\begin{equation}
\sum_{J_z=-1,0,1}\left[ O^{0,s_{NL}}_{1,2,1,J_z,0,0}({\bf
n})\right]^{\dagger}O^{0,s_{NL}}_{1,0,1,J_z,0,0}({\bf
n})+\sum_{J_z=-1,0,1}\left[ O^{0,s_{NL}}_{1,0,1,J_z,0,0}({\bf
n})\right]^{\dagger}O^{0,s_{NL}}_{1,2,1,J_z,0,0}({\bf
n}). 
\end{equation}
\subsection{Order $Q^4$}
At order $Q^4$ we have the next-to-lowest radial excitations of the $S$-wave
spin
singlet $V_{4,^1S_0,1}({\bf n})$, 
\begin{equation}
\sum_{I_z=-1,0,1}\left[ O^{2,s_{NL}}_{0,0,0,0,1,I_z}({\bf n})\right]^{\dagger}O^{2,s_{NL}}_{0,0,0,0,1,I_z}({\bf
n}),  
\end{equation}
and $V_{4,^1S_0,2}({\bf n}),$
\begin{equation}
\sum_{I_z=-1,0,1}\left[ O^{4,s_{NL}}_{0,0,0,0,1,I_z}({\bf n})\right]^{\dagger}O^{0,s_{NL}}_{0,0,0,0,1,I_z}({\bf
n}) + \sum_{I_z=-1,0,1}\left[ O^{0,s_{NL}}_{0,0,0,0,1,I_z}({\bf n})\right]^{\dagger}O^{4,s_{NL}}_{0,0,0,0,1,I_z}({\bf
n}),
\end{equation}
and the next-to-lowest radial excitations of the $S$-wave spin triplet
$V_{4,^3S_1,1}({\bf n}),$
\begin{equation}
\sum_{J_z=-1,0,1}\left[ O^{2,s_{NL}}_{1,0,1,J_z,0,0}({\bf n})\right]^{\dagger}O^{2,s_{NL}}_{1,0,1,J_z,0,0}({\bf
n})
\end{equation}
and $V_{4,^3S_1,2}({\bf n}),$ 
\begin{equation}
\sum_{J_z=-1,0,1}\left[ O^{4,s_{NL}}_{1,0,1,J_z,0,0}({\bf n})\right]^{\dagger}O^{0,s_{NL}}_{1,0,1,J_z,0,0}({\bf
n}) + \sum_{J_z=-1,0,1}\left[ O^{0,s_{NL}}_{1,0,1,J_z,0,0}({\bf n})\right]^{\dagger}O^{4,s_{NL}}_{1,0,1,J_z,0,0}({\bf
n}).
\end{equation}
If we apply the on-shell equivalence condition that the magnitude of the
outgoing relative momentum equals the magnitude of the incoming relative momentum,
then $V_{4,^1S_0,1}$
and $V_{4,^1S_0,2}$ 
are equivalent and also $V_{4,^3S_1,1}$ and $V_{4,^3S_1,2}$
are equivalent. In this work we make the choice of setting  the coefficients of 
$V_{4,^1S_0,2}$ and $V_{4,^3S_1,2}$ to 0.

At order $Q^4$ we have the  first radial excitations of the $^1P_1$ interaction
$V_{4,^1P_1}({\bf n}),$
\begin{equation}
\sum_{J_z=-1,0,1}\left[ O^{2,s_{NL}}_{0,1,1,J_z,0,0}({\bf n})\right]^{\dagger}O^{0,s_{NL}}_{0,1,1,J_z,0,0}({\bf
n})+\sum_{J_z=-1,0,1}\left[ O^{0,s_{NL}}_{0,1,1,J_z,0,0}({\bf n})\right]^{\dagger}O^{2,s_{NL}}_{0,1,1,J_z,0,0}({\bf
n}), 
\end{equation}
the $^3P_0$ interaction $V_{4,^3P_0}({\bf n}),$
\begin{equation}
\sum_{I_z=-1,0,1}\left[ O^{2,s_{NL}}_{1,1,0,0,1,I_z}({\bf n})\right]^{\dagger}O^{0,s_{NL}}_{1,1,0,0,1,I_z}({\bf
n})+\sum_{I_z=-1,0,1}\left[ O^{0,s_{NL}}_{1,1,0,0,1,I_z}({\bf n})\right]^{\dagger}O^{2,s_{NL}}_{1,1,0,0,1,I_z}({\bf
n}), 
\end{equation}
the $^3P_1$ interaction $V_{4,^3P_1}({\bf n}),$
\begin{equation}
\sum_{I_z=-1,0,1}\sum_{J_z=-1,0,1}\left[ O^{2,s_{NL}}_{1,1,1,J_z,1,I_z}({\bf
n})\right]^{\dagger}O^{0,s_{NL}}_{1,1,1,J_z,1,I_z}({\bf
n}) \\ +\sum_{I_z=-1,0,1}\sum_{J_z=-1,0,1}\left[ O^{0,s_{NL}}_{1,1,1,J_z,1,I_z}({\bf
n})\right]^{\dagger}O^{2,s_{NL}}_{1,1,1,J_z,1,I_z}({\bf
n}), 
\end{equation}
and the $^3P_2$ interaction $V_{4,^3P_2}({\bf n}),$
\begin{equation}
\sum_{I_z=-1,0,1}\sum_{J_z=-2,...2}\left[ O^{2,s_{NL}}_{1,1,2,J_z,1,I_z}({\bf
n})\right]^{\dagger}O^{0,s_{NL}}_{1,1,2,J_z,1,I_z}({\bf
n})+\sum_{I_z=-1,0,1}\sum_{J_z=-2,...2}\left[ O^{0,s_{NL}}_{1,1,2,J_z,1,I_z}({\bf
n})\right]^{\dagger}O^{2,s_{NL}}_{1,1,2,J_z,1,I_z}({\bf
n}). 
\end{equation}
At order $Q^4$ we also have the first radial excitations of the $S-D$ mixing term
$V_{4,SD,1}({\bf n}),$
\begin{equation}
\sum_{J_z=-1,0,1}\left[ O^{2,s_{NL}}_{1,2,1,J_z,0,0}({\bf
n})\right]^{\dagger}O^{0,s_{NL}}_{1,0,1,J_z,0,0}({\bf
n})+\sum_{J_z=-1,0,1}\left[ O^{0,s_{NL}}_{1,0,1,J_z,0,0}({\bf
n})\right]^{\dagger}O^{2,s_{NL}}_{1,2,1,J_z,0,0}({\bf
n}),
\end{equation}
and $V_{4,SD,2}({\bf n}),$
\begin{equation}
\sum_{J_z=-1,0,1}\left[ O^{0,s_{NL}}_{1,2,1,J_z,0,0}({\bf
n})\right]^{\dagger}O^{2,s_{NL}}_{1,0,1,J_z,0,0}({\bf
n})+\sum_{J_z=-1,0,1}\left[ O^{2,s_{NL}}_{1,0,1,J_z,0,0}({\bf
n})\right]^{\dagger}O^{0,s_{NL}}_{1,2,1,J_z,0,0}({\bf
n}). 
\end{equation}
If we apply the on-shell equivalence condition, then $V_{4,SD,1}$ and $V_{4,SD,2}$
are equivalent.
 In this work we make the choice of setting  the coefficient
of $V_{4,SD,1}$ to 0.

At order $Q^4$ we have the $^1D_2$ interaction
$V_{4,^1D_2}({\bf n}),$
\begin{equation}
\sum_{I_z=-1,0,1}\sum_{J_z=-2,...2}\left[ O^{0,s_{NL}}_{0,2,2,J_z,1,I_z}({\bf
n})\right]^{\dagger}O^{0,s_{NL}}_{0,2,2,J_z,1,I_z}({\bf
n}),
\end{equation}
the $^3D_1$ interaction $V_{4,^3D_1}({\bf n}),$
\begin{equation}
\sum_{J_z=-1,0,1}\left[ O^{0,s_{NL}}_{1,2,1,J_z,0,0}({\bf
n})\right]^{\dagger}O^{0,s_{NL}}_{1,2,1,J_z,0,0}({\bf
n}),
\end{equation}
the $^3D_2$ interaction $V_{4,^3D_2}({\bf n}),$
\begin{equation}
\sum_{J_z=-2,...2}\left[ O^{0,s_{NL}}_{1,2,2,J_z,0,0}({\bf
n})\right]^{\dagger}O^{0,s_{NL}}_{1,2,2,J_z,0,0}({\bf
n}),
\end{equation}
and the $^3D_3$ interaction $V_{4,^3D_3}({\bf n}),$
\begin{equation}
\sum_{J_z=-3,...3}\left[ O^{0,s_{NL}}_{1,2,3,J_z,0,0}({\bf
n})\right]^{\dagger}O^{0,s_{NL}}_{1,2,3,J_z,0,0}({\bf
n}).
\end{equation}
At order $Q^4$ we also have the $P-F$ mixing term $V_{4,PF}({\bf n})$,
\begin{equation}
\sum_{I_z=-1,0,1}\sum_{J_z=-2,...2}\left[ O^{0,s_{NL}}_{1,3,2,J_z,0,0}({\bf
n})\right]^{\dagger}O^{0,s_{NL}}_{1,1,2,J_z,0,0}({\bf
n})+\sum_{I_z=-1,0,1}\sum_{J_z=-2,...2}\left[ O^{0,s_{NL}}_{1,1,2,J_z,0,0}({\bf
n})\right]^{\dagger}O^{0,s_{NL}}_{1,3,2,J_z,0,0}({\bf
n}). 
\end{equation}  

\subsection{Isospin-breaking short-range interactions}
We also include additional isospin-breaking $^1S_0$ contact interactions
for proton-proton scattering ($I_z=1$) and neutron-neutron scattering ($I_z=-1$).
These are not relevant for neutron-proton scattering,
but we nevertheless discuss the interactions for completeness.
We define the two isospin-breaking interactions $V^{I_z=1}_{0,^1S_0}({\bf n}),$ 
\begin{equation}
\left[ O^{0,s_{NL}}_{0,0,0,0,1,I_z=1}({\bf n})\right]^{\dagger}O^{0,s_{NL}}_{0,0,0,0,1,I_z=1}({\bf
n})  
\end{equation}
and $V^{I_z=-1}_{0,^1S_0}({\bf
n})$, 
\begin{equation}
\left[ O^{0,s_{NL}}_{0,0,0,0,1,I_z=-1}({\bf n})\right]^{\dagger}O^{0,s_{NL}}_{0,0,0,0,1,I_z=-1}({\bf
n}).  
\end{equation}
In terms of counting momenta, these are order $Q^0$. However, they are suppressed
by the small size of the isospin-breaking coefficient.  Following our previous analyses,
we count this correction as order $Q^2$.  We do not consider higher-order isospin breaking 
terms in this work, but they will be included in future studies.

\section{Long-range interactions}\label{long:range}
\subsection{One-pion exchange}
The one-pion exchange interaction $V_{\rm OPE}$ has the form
\begin{align}
V_{\rm OPE}=-\frac{g_A^2}{8F^2_{\pi}}\sum_{{\bf n',n},S',S,I}
:\rho_{S',I\rm }({\bf n'})f_{S'S}({\bf n'}-{\bf n})\rho_{S,I}({\bf n}):,
\end{align}
where $f_{S'S}$ is defined as
\begin{align}
f_{S'S}({\bf n'}{\bf -n})=\frac{1}{L^3}\sum_{\bf q}\frac{Q(q_{S'})Q(q_{S})\exp[-i{\bf q}\cdot({\bf
n'}-{\bf n})-b_{\pi}({\bf q}^2+M^2_{\pi})]}{{\bf q}^2 + M_{\pi}^2},\label{OPE}
\end{align}
where $L$ is the length of the cubic periodic box and each lattice momentum component $q_S$ is an integer multiplied by $2\pi/L$.
The function $Q(q_S)$ is defined as 
\begin{equation}
Q(q_S) = \frac{3}{2}\sin(q_S)-\frac{3}{10}\sin(2q_S)+\frac{1}{30}\sin(3q_S), \label{momentum:q}
\end{equation}
which equals $q_S$ up to a correction of order $q_S^7$.
The parameter
$b_{\pi}$ is included to remove short-distance lattice artifacts
in the one-pion exchange interaction.  In the present calculation, we set  $b_\pi = 0.25$ in lattice units.  We have used the combination ${\bf
q}^2+M^2_{\pi}$ in the exponential as suggested in recent work  \cite{Reinert:2017usi} as a momentum-space regulator
that does not affect the long-distance behavior.  At leading order we take the pion mass to be the mass of the neutral pion,
$M_{\pi} = M^0_{\pi} =M_{{\pi},I=3}$.

\subsection{Two-pion exchange}

The cutoff momentum arising from the lattice regularization is $\Lambda_{\rm latt} = \pi/a$, with $a$ the spatial lattice spacing. For coarse lattice spacings such  as
$a = 1.97$ and $a =1.64$ fm, the corresponding lattice cutoffs are  $314$ and $377~{\rm MeV}$, respectively. For  momenta lying below these cutoff scales, 
the two-pion-exchange potential (TPEP) can be expanded in powers of ${\bf q}^2/(4\pi^2)$, resulting in operators that are exactly the same as our short-range contact terms. We conclude that TPEP at coarse lattice spacings can be replaced by retuning 
the low-energy constants (LECs) for these contact terms \cite{Borasoy:2007vi}. For these two coarse lattice spacings, the TPEP does not have observable effects 
but only changes the LECs. 

For the two smaller lattice spacings, $a = 1.32~{\rm fm}$ and $a = 0.99~{\rm fm}$, however, higher momenta can be reached and the structure of the two-pion-exchange potential can be resolved. In these cases we include the TPEP explicitly. According to the power counting of chiral 
EFT, the TPEP first appears at order $O(Q^2)$ or NLO, the subleading TPEP appears at order $O(Q^3)$ or N$^2$LO, and so on \cite{Weinberg:1990rz}. 
Similarly  to what we do to the one-pion exchange potential,  we also regularize the TPEP by a Gaussian form factor in  momentum space, 
 
\begin{eqnarray}
F(\bm{q})=\exp\left[-b_\pi(\bm{q}^2 + M_{\pi}^2)\right] = \exp\left[-\frac{\bm{q}^2 + M_{\pi}^2}{\Lambda^2}\right],
\end{eqnarray}
where $\Lambda = 1/{\sqrt{b_\pi}}$~\cite{Reinert:2017usi}. 
In the present calculation, we set  $b_\pi = 0.25$ lattice units, which equates to  $\Lambda = 300~{\rm MeV}$ for $a = 1.32~{\rm fm}$ 
and $\Lambda = 400~{\rm MeV}$ for $a = 0.99~{\rm fm}$. 
In this work, the relativistic corrections stemming from the 
$1/{m_N^2}$ corrections to the one-pion exchange potential and $1/m_N$ correction to the TPEP at order $O(Q^4)$  are not taken into account. 

The TPEP up to order $O(Q^4)$ or N$^3$LO is completely local and can be written in the form 
\begin{eqnarray}
V_{\rm TPEP} &=& V_{\mathrm{TPEP}}^{Q^2} + V_{\mathrm{TPEP}}^{Q^3} + V_{\mathrm{TPEP}}^{Q^4}  \nonumber \\
&=& \frac{1}{2}\sum_{{\bf n}, {\bf n}^\prime} :\rho({\bf n}) V_C({\bf n} - {\bf n}^\prime) \rho({\bf n}^\prime):  
+ \frac{1}{2}\sum_I \sum_{{\bf n}, {\bf, n}^\prime} :\rho_I({\bf n}) W_C({\bf n} - {\bf n}^\prime) \rho_I({\bf n}^\prime):   \nonumber\\
&~& + \frac{1}{2}\sum_S\sum_{{\bf n}, {\bf n}^\prime} :\rho_S({\bf n}) V_\sigma({\bf n} - {\bf n}^\prime) \rho_S({\bf n}^\prime):
 + \frac{1}{2}\sum_{S,I}\sum_{{\bf n}, {\bf n}^\prime} :\rho_{S,I}({\bf n}) W_{\sigma}({\bf n} - {\bf n}^\prime) \rho_{S,I}({\bf n}^\prime):  \\
&~& + \frac{1}{2}\sum_{S_1,S_2}\sum_{{\bf n}, {\bf n}^\prime}:\rho_{S_1}({\bf n}) (V_T)_{S_1,S_2}({\bf n} - {\bf n}^\prime) \rho_{S_2}({\bf n}^\prime): 
 + \frac{1}{2}\sum_{S_1,S_2,I}\sum_{{\bf n}, {\bf n}^\prime} :\rho_{S_1,I}({\bf n}) (W_T)_{S_1,S_2}({\bf n} - {\bf n}^\prime) \rho_{S_2,I}({\bf n}^\prime):  \nonumber
\end{eqnarray}
where $V_{C/\sigma}$,  $(V_T)_{S_1, S_2}$, $W_{C/\sigma}$, and $(W_T)_{S_1,S_2}$  are scalar functions in the coordinate space, and
\begin{eqnarray}
V_{C/\sigma}({\bf n}  - {\bf n}^\prime) = \frac{1}{L^3} \sum_{\bf q} \exp \left[-i{\bf q}\cdot({\bf n} - {\bf n}^\prime)\right] V_{C/\sigma}({\bf q}) F({\bf q}), 
\end{eqnarray}
\begin{eqnarray}
W_{C/\sigma}({\bf n}  - {\bf n}^\prime) = \frac{1}{L^3} \sum_{\bf q} \exp
\left[-i{\bf q}\cdot({\bf n} - {\bf n}^\prime)\right] W_{C/\sigma}({\bf q})
F({\bf q}), 
\end{eqnarray}

\begin{eqnarray}
(V_T)_{S_1,S_2}({\bf n}  - {\bf n}^\prime) = \frac{1}{L^3} \sum_{\bf q} \exp\left[-i{\bf q}\cdot({\bf n} - {\bf n}^\prime)\right] (V_T)_{S_1, S_2}({\bf q}) F({\bf q})Q(q_{S_1})Q(q_{S_2}),
\end{eqnarray}
\begin{eqnarray}
(W_T)_{S_1,S_2}({\bf n}  - {\bf n}^\prime) = \frac{1}{L^3} \sum_{\bf q} \exp\left[-i{\bf
q}\cdot({\bf n} - {\bf n}^\prime)\right] (W_T)_{S_1, S_2}({\bf q}) F({\bf
q})Q(q_{S_1})Q(q_{S_2}).
\end{eqnarray}
 The definitions for the functions $V_{C/\sigma}({\bf q})$, $(W_T)_{C/\sigma}({\bf q})$, 
$(V_T)_{S_1, S_2}({\bf q}),$ and $(W_T)_{S_1, S_2}({\bf q})$  are given in Refs. \cite{Kaiser:2001pc,Epelbaum:2014efa,Entem:2014msa,Reinert:2017usi,Epelbaum:2004fk}.

\subsection{Coulomb and long-range strong isospin breaking}

The Coulomb interaction will not be relevant for neutron-proton scattering,
but we nevertheless discuss it here for completeness.
The Coulomb interaction can be written as
\begin{align}
V_{\rm Coulomb}=-\frac{\alpha_{\rm EM}}{2_{}}\sum_{{\bf n',n}}
:\frac{1}{4}[\rho({\bf n'})+\rho_{I=3\rm }({\bf n'})]\frac{1}{d({\bf
n'}-{\bf n})}[\rho({\bf n})+\rho_{I=3\rm }({\bf n})]:,
\end{align}
where $d({\bf
n'}-{\bf n})$ is the shortest length of ${\bf
n'}-{\bf n}$ as measured on the periodic lattice, and we define the value
of $d$ at the
origin to be $1/2$.
Our notation $\rho_{I=3\rm }$ refers to the $I=3$ isospin component of $\rho_{I}$.

The long-range isospin-breaking correction, due to differences in the charged and neutral pion mass in one-pion exchange, has the form
\begin{align}
V^{\rm IB}_{\rm OPE}=-\frac{g_A^2}{8f^2_{\pi}}\sum_{{\bf n',n},S',S,I}
:\rho_{S',I\rm }({\bf n'})f^{\rm IB}_{S'SI}({\bf n'}-{\bf n})\rho_{S,I}({\bf n}):,
\end{align}
where $f^{\rm IB}_{S'SI}$ is defined as
\begin{align}
f^{\rm IB}_{S'SI}({\bf n'}{\bf -n})=\frac{1}{L^3}\sum_{\bf q}\frac{Q(q_{S'})Q(q_{S})\exp[-i{\bf q}\cdot({\bf
n'}-{\bf n})-b_{\pi}({\bf q}^2+M^2_{\pi,I})]}{{\bf q}^2 + M_{\pi,I}^2}-f^{\rm }_{S'S}({\bf n'}{\bf -n}),\label{OPE2}
\end{align}
where $M_{{\pi},1} = M_{{\pi},2}=M^+_{\pi}=M^-_{\pi}$ and $M_{{\pi},3} = M^0_{{\pi}}$.
As in previous analyses
we count this correction as order $Q^2$.

\section{Galilean Invariance Restoration (GIR)}

Galilean invariance is the statement that the laws of Newtonian physics for a non-relativistic system are independent of the velocity of the center of mass. In a lattice regularized system, however, the effect of the cutoff is different in moving frames, and this leads to the breaking of Galilean invariance~\cite{Lee:2007jd}. There is also some breaking of Galilean invariance caused by the nonlocal smearing parameter $s_{\rm NL}$ that we use in the construction of our interactions.  This arises from the residual dependence of the interactions on the velocity of the center of mass.  Fortunately, in many cases of interest these two Galilean invariance breaking effects have the tendency to partially cancel.  

In order to restore Galilean invariance in the two-nucleon system, we include the two-nucleon nearest-neighbor hopping operator, 
\begin{equation}
V_{\rm GIR} = V^0_{\rm GIR} + V^1_{\rm GIR},
\end{equation}
where
\begin{align}
V^0_{\rm GIR} =C_{\rm GIR} \sum_{{\bf n},i,j,i',j'} a^{\dagger}_{i,j}({\bf n})a^{\dagger}_{i',j'}({\bf n})a_{i',j'}({\bf
n})a_{i,j}({\bf
n})
\end{align}
and 
\begin{align}
V^1_{\rm GIR} = -\frac{C_{\rm GIR}}{6}\sum_{{\bf n},i,j,i',j'}\sum_{\bf n'} 
a^{\dagger}_{i,j}({\bf n}+{\bf n'})
a^{\dagger}_{i',j'}({\bf n}+{\bf n'})
a_{i',j'}({\bf n})a_{i,j}({\bf n}).
\end{align}
Let us write $\lvert {\bf P}_{\rm tot}\rangle$ as a two-body bound-state wave function with total momentum ${\bf P}_{\rm tot}$.  We note that 
\begin{equation}
\langle{{\bf P}_{\rm tot}} \vert V^0_{\rm GIR} \vert {\bf P}_{\rm tot} \rangle\\ \end{equation}
is independent of ${\bf P}_{\rm tot}$, and so we have 
\begin{equation}
\langle{{\bf P}_{\rm tot}} \vert V^0_{\rm GIR} \vert {\bf P}_{\rm tot} \rangle =\langle {\bf 0} \vert V^0_{\rm GIR} \vert {\bf 0} \rangle,
\end{equation}
where $\lvert {\bf 0}\rangle$ is the two-body bound-state
wave function with zero total momentum.  Furthermore,
\begin{align}
\langle{ {\bf P}_{\rm tot}} & \vert V^1_{\rm GIR} \vert {\bf P}_{\rm tot} \rangle=
-\frac{1}{6} \langle {\bf 0} \vert V^0_{\rm GIR} \vert {\bf 0} 
\rangle\sum_{\bf n'}
e^{-i {\bf P}_{\rm tot}\cdot \vec{\bf n}'} \nonumber \\
& = \left[- \frac{1}{3} \cos(P_{{\rm tot},1}) -\frac{1}{3} \cos(P_{{\rm tot},2}) - \frac{1}{3} \cos(P_{{\rm tot},3}) \right]
\langle {\bf 0} \vert V^0_{\rm GIR} \vert {\bf 0} \rangle.
\end{align}
Therefore
\begin{align}
\langle{ {\bf P}_{\rm tot}} & \vert V_{\rm GIR} \vert {\bf P}_{\rm tot}
\rangle
 \nonumber \\
& = [1 - \frac{1}{3} \cos(P_{{\rm tot},1}) -\frac{1}{3} \cos(P_{{\rm tot},2})
- \frac{1}{3} \cos(P_{{\rm tot},3})]
\langle {\bf 0} \vert V^0_{\rm GIR} \vert {\bf 0} \rangle.
\end{align}
In this manner we can restore Galilean invariance up to order $Q^2$ by tuning the coefficient of $ V_{\rm GIR}$ according to the dispersion relation of the deuteron 
and the $^1S_0$ ground state at finite volume.  The deuteron dispersion relation is far more useful 
for this purpose, however, since the $^1S_0$ ground state is a continuum state that shows negligible Galilean invariance breaking in its dispersion relation. This is true for all continuum states, and this is why the amount of Galilean invariance breaking seen in the higher partial waves are also negligible.  We will consider Galilean invariance breaking effects beyond order $Q^2$ in future work.  

\begin{figure}
\centering
\begin{tabular}{cc}
\includegraphics[width=0.45\textwidth]{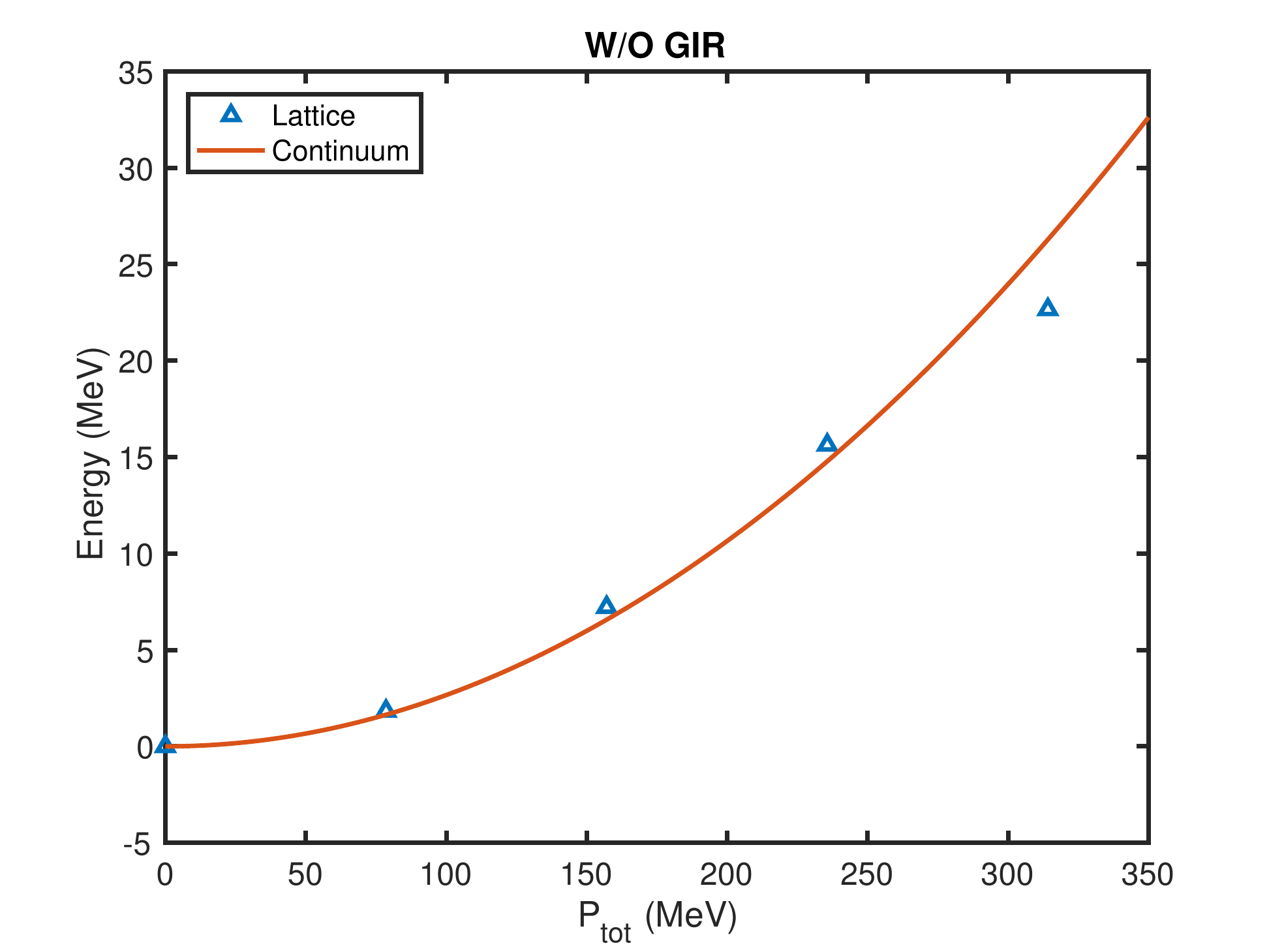} & 
\includegraphics[width=0.45\textwidth]{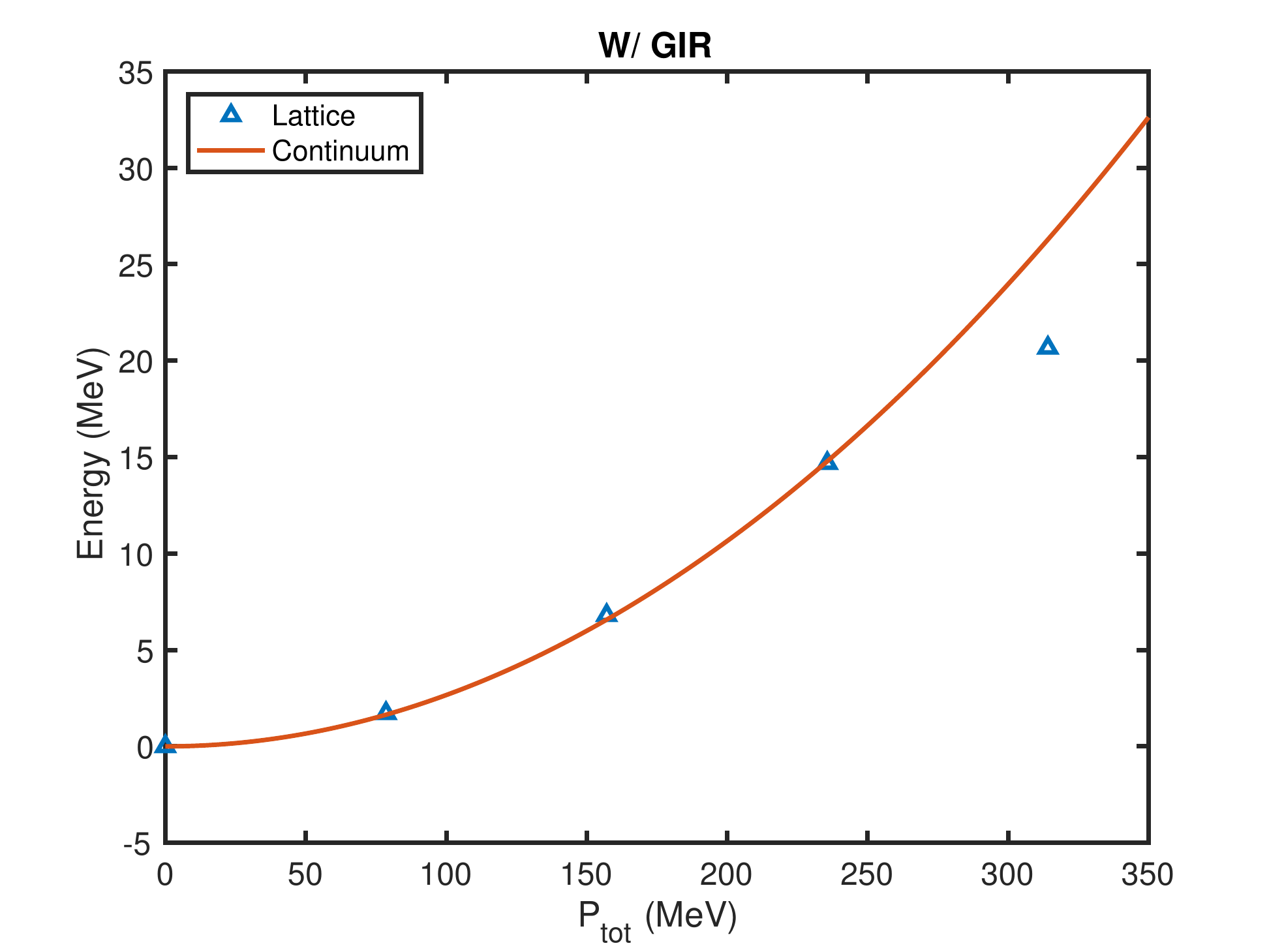}\\
\end{tabular}
\caption{(Color online) Dispersion relation of the deuteron. Left: Without  any Galilean invariance restoration (GIR). 
Right: With  GIR as provided by the operator $V_{\rm GIR}$ with coefficient $C_{\mathrm{GIR}} = -0.0658$.}\label{deuteronGIR}
\end{figure}

As an example of how to determine the 
Galilean invariance restoration operator coefficient, we show the results for lattice spacing $a = 1.97~{\rm fm}$ in Fig.~(\ref{deuteronGIR}). The left and right panels 
are the deuteron dispersion relations before and after including the Galilean invariance restoration operator respectively. We see that the amount of correction is relatively small.  The fitted coefficient for $V_{\rm GIR}$ 
is found to be $C_{\rm GIR} = -0.0658$.
 The amount of Galilean invariance breaking
is somewhat smaller than this for the smaller lattice spacings.  

\section{Scattering on the lattice}
In order to calculate  the scattering phase shifts and mixing angles, we first construct radial wave functions through the spherical harmonics 
with quantum numbers $(L, L_z)$ \cite{Lu:2015riz, Elhatisari:2016hby}, 
\begin{eqnarray}
\left|r\right>^{L, L_z} = \sum_{\bf r^\prime} Y_{L, L_z}({\hat r}^\prime) \delta_{|{\bf r}^\prime| = r} \left|{\bf r}^\prime\right>,
\end{eqnarray}
where ${\bf r}^\prime$ runs over all lattice grid points having the same radial lattice distance.  We group 
together data into a large number of radial bins so that in each bin, $r- \delta r/2<|{\bf r}^\prime|<r+ \delta r/2$, with a very small width parameter
$\delta r$.  Using this definition of  the radial wave function, the Hamiltonian matrix  over a three-dimensional lattice can be reduced 
to a one-dimensional radial Hamiltonian, $H_{{\bf r}, {\bf r}^\prime} \rightarrow H_{r, r^\prime}$. 

We follow the method described in
Ref.~\cite{Lu:2015riz}, which uses an auxiliary radial potential.  We extract the phase shifts as well as the mixing angles from the radial wave functions in the 
region where the NN force  and auxiliary potentials are vanishing.  In this range, the wave function has the form,
\begin{eqnarray}
A_L h_L^-(kr) - B_Lh_L^+(kr),  
\end{eqnarray}
where $h_L^-(kr)$ and $h_L^+(kr)$ are the spherical Bessel functions, $k = \sqrt{2\mu E}$, $\mu$ is the reduced mass, and $E$ is the energy.  
 The scattering coefficients $A_L$ and $B_L$ satisfy the 
relations, 
\begin{eqnarray}
B_L = S_LA_L,  \label{eq:ph}
\end{eqnarray}
where $S_L = \exp\left(2i \delta_L\right)$ is the $S$-matrix and $\delta_L$ is the phase shift. The phase shift  is determined by setting 
\begin{eqnarray}
\delta_L = \frac{1}{2i}\log\left(\frac{B_L}{A_L}\right).
\end{eqnarray}
In the case of the coupled channels with $J > 0$, both of the coupled partial waves, $L = J -1$ and $L= J + 1$, satisfy Eq. (\ref{eq:ph}), 
and the $S$-matrix couples the two channels together. Throughout this work  we adopt the so-called  Stapp parametrization of
 the phase shifts and mixing angles for the coupled channels \cite{Stapp:1956mz}, 
\begin{eqnarray}
S = \left[
\begin{array}{cc}
\cos(2\epsilon) \exp \left(2i\delta_{J-1}^{1J} \right)   &    i\sin(2\epsilon) \exp \left(i \delta_{J-1}^{1J} + i \delta_{J+1}^{1J} \right) \\
i \sin{2\epsilon} \exp \left(i\delta_{J-1}^{1J} + i \delta_{J+1}^{1J} \right)   &  \cos(2\epsilon) \exp \left( 2i \delta_{J+1}^{1J} \right) \\
\end{array}
\right].
\end{eqnarray}

\section{Results for the Neutron-proton phase shifts}\label{compare}
Different lattice spacings introduce different lattice artifacts. We make calculations using four different lattice spacings,  $ a = 1.97$, 
$1.64$, $1.32$ and $0.99~{\rm fm}$ to study the lattice spacing effects. We choose these values because the corresponding 
lattice momentum cutoffs, $\Lambda_{\rm latt} = \pi/a$, remain below the estimated breakdown scale of chiral effective field theory and  the 
order-by-order convergence has been demonstrated to be favorable in few-body and many-body calculations.
As noted in discussion above, we do not include the TPEP for the two coarse lattice spacings, $a = 1.97~{\rm fm}$ and $1.64~{\rm fm}$. For the two smaller lattice
spacings, $a = 1.32~{\rm fm}$ and $0.99~{\rm fm}$, we present results both with and without the TPEP in order to discern the effect of the
TPEP.

In previous lattice studies we had to contend with interactions that had an effect in all channels. With these new lattice interactions this problem is now completely solved.  We need only
to consider the interactions that participate in a given channel, and our labeling of the operators makes clear which channels these are. We determine the LECs by 
reproducing the neutron-proton scattering phase shifts and mixing angles of the Nijmegen partial wave analysis (NPWA) \cite{Stoks:1993tb}. 
Since the NPWA provides only the statistical errors, and not the systematic errors, we use the procedure described in Ref.~\cite{Epelbaum:2014efa} to account for the systematic uncertainties. Specifically, we use
\begin{eqnarray}
\Delta_X = \max\left(\Delta_X^{\mathrm{NPWA}}, |\delta_X^{\mathrm{NijmI}} - \delta_X^{\mathrm{NPWA}}|, 
| \delta_X^{\mathrm{NijmII}} - \delta_X^{\mathrm{NPWA}} |, | \delta_X^{\mathrm{Reid93}} - \delta_X^{\mathrm{NPWA}} | \right), 
\end{eqnarray} 
where $\delta_X^i$  are the phase shifts  (mixing angles )
in channel $X$ based on different NPWA potentials, while $\Delta_X^{\mathrm{NPWA}}$ are the statistical errors of the phase shifts 
(mixing angles) of the NPWA. 

For the coupled channel, ${}^3S_1 - {}^3D_1$, we define the  $\chi^2$ as
\begin{eqnarray}
\chi^2 = \sum_i\frac{(\delta^{\rm Latt}_i - \delta_i^{\rm NPWA})^2}{\Delta_i^2}  + \frac{(E_b^{\mathrm{Latt}}-E_b^{\mathrm{Exp}})^2}{\Delta E_{\mathrm{Exp}}^2},
\end{eqnarray}
with the deuteron binding energy  $E_b^{\rm Exp} = 2.224575$ MeV and corresponding error $\Delta E_{\rm Exp} = 9\times 10^{-6}$ MeV.  
For the other channels, we define $\chi^2$ as 
\begin{eqnarray}
\chi^2 = \sum_i\frac{(\delta^{\mathrm{Latt}} - \delta_i^{\mathrm{NPWA}})^2}{\Delta_i^2}.
\end{eqnarray} 

In our fits we choose energy ranges that are appropriate for the chiral order and lattice spacing used. Specifically, 
for the coarser lattice spacings,  $a = 1.97$  and $1.64~{\rm fm}$, we take the energy range  $E_{\rm lab} \le 50~{\rm MeV}$ for the 
LO, NLO/N$^2$LO, and N$^3$LO fits. In those cases we use five points, $E_{\rm lab}  = 1$, 5, 10, 25, and 50~MeV,  to compute the corresponding $\chi^2$. For the fits with the smaller lattice spacings, $a = 1.32$ and $0.99$~fm, we take the energy range $E_{\rm lab} \le 50$~MeV 
for the LO, NLO and N$^2$LO fits, and $E_{\rm lab} \le 100$~MeV for the N$^3$LO fits. Thus we determine the 
$\chi^2$ for the N$^3$LO fits using six points, $E_{\rm lab} = 1$, 5, 10, 25, 50 and 100~MeV.  The LECs determined by the N$^3$LO fits 
are listed in Table \ref{LECs} for each of the lattice 
spacings. 
\begin{table}
\renewcommand{\arraystretch}{1.5}
\centering
\caption{Low-energy constants determined by N$^3$LO fits using lattice spacings, $a = 1.97~{\rm fm}$ , $1.64~{\rm fm}$,
 $1.32~{\rm fm}$ and $0.99~{\rm fm}$.
For calculations using $a = 1.32~{\rm fm}$ and $0.99~{\rm fm}$, full NN interactions are used. All LECs are given in lattice units.}\label{LECs}
\begin{tabular*}{0.95\textwidth}{@{\extracolsep{\fill}} l|rrrr}
\hline\hline
      LECs                    &  $ a = 1.97$ fm ~~~    &  $a = 1.64$ fm ~~~   &  $a = 1.32$ fm~~~~   &      $a = 0.99$ fm ~~~~  \\   
  \hline
 $C_{0,{}^1S_0}$  & $ 0.1050\pm0.0006$ &  $ 0.0879\pm 0.0004$ & $0.0833\pm 0.0010$ & $ 0.0860\pm0.0004$  \\
 $C_{0,{}^3S_1}$  & $ 0.0256\pm0.0056$ & $ 0.0322\pm0.0031$   & $ 0.0455\pm0.0289$ & $ 0.0520\pm0.0006$ \\
 \hline
 $C_{2,{}^1S_0}$            & $ 0.0217\pm0.0002$ & $ 0.0242\pm0.0002$  & $0.0271\pm 0.0007$ & $0.0256\pm0.0005$ \\
 $C_{2,{}^3S_1}$            & $ 0.0267\pm0.0020$ & $ 0.0280\pm0.0014$  & $ 0.0310\pm0.0179$ &$0.0263\pm0.0005$  \\
 $C_{2,SD}$                   & $-0.0605\pm0.0041$ & $-0.0421\pm0.0047$ & $-0.0291\pm0.0137$ &$-0.0089\pm 0.0021$ \\
 $C_{2,{}^1P_1}$            & $ 0.1930\pm 0.0012$ & $0.1758\pm0.0013$ & $0.1469 \pm0.0003 $ &$0.1321\pm 0.0002$ \\
 $C_{2,{}^3P_0}$            & $-0.0084\pm0.0004$ & $0.0190\pm 0.0007$ & $0.0495\pm0.0004 $ &$ 0.0940\pm0.0003$ \\ 
 $C_{2,{}^3P_1}$            & $ 0.1332\pm0.0013$ & $ 0.1217 \pm0.0007$& $ 0.1186\pm0.0034$ &$0.1300\pm0.0007$ \\
 $C_{2,{}^3P_2}$            & $ 0.0441\pm0.0001$ & $0.0461 \pm0.0018$ & $0.0584\pm 0.0021$ &$0.0665\pm0.0002$ \\
 \hline
 $C_{4,{}^1S_0}$            & $0.0073\pm 0.0001$ & $0.0081\pm0.0001$ & $0.0108\pm 0.0005$ & $0.0148\pm0.0006$\\
 $C_{4,{}^3S_1}$            & $0.0079\pm 0.0007$ & $0.0081\pm0.0006$ & $ 0.0119\pm0.0109$ & $0.0153\pm0.0004$ \\ 
 $C_{4,SD}$                   & $0.0005\pm 0.0003$ & $-0.0011\pm0.0006$ &$-0.0026\pm0.0029$& $-0.0098\pm0.0011$ \\ 
 $C_{4,{}^1P_1}$            & $-0.0004\pm0.0006$ & $ -0.0057\pm0.0006$ &$ -0.0104\pm 0.0001$&$-0.0105\pm0.0002$ \\
 $C_{4,{}^3P_0}$            & $-0.0001\pm0.0002$ & $ -0.0006\pm0.0005$ &$-0.0024\pm 0.0001$&$-0.0022\pm 0.0007$\\ 
 $C_{4,{}^3P_1}$            & $-0.0006\pm0.0004$ & $ -0.0004\pm0.0004$ & $-0.0019\pm0.0013$&$0.0063\pm0.0013$\\ 
 $C_{4,{}^3P_2}$            & $0.0080\pm 0.0002$ & $ 0.0090\pm 0.0012$ &$0.0105\pm 0.0008$&$0.0078\pm0.0003$ \\ 
 $C_{4,PF}$                    & $0.0072\pm 0.0002$  & $ 0.0041\pm 0.0011$ & $0.0017\pm 0.0002$&$0.0026\pm0.0002$ \\ 
 $C_{4,{}^1D_2}$            & $0.0105\pm 0.0006$  & $ 0.0088\pm 0.0005$ &$0.0136\pm0.0001$ &$0.0190\pm0.0050$ \\
 $C_{4,{}^3D_1}$            & $ 0.0327\pm 0.0023$ &$0.0319\pm0.0039 $  &$ 0.0318\pm0.0134$ &$0.0720\pm0.0122$ \\ 
 $C_{4,{}^3D_2}$            & $-0.032\pm 0.0017$ & $-0.0324\pm0.0019$  &$-0.0187\pm0.0022$ &$-0.0005\pm0.0014$ \\ 
 $C_{4,{}^3D_3}$            & $0.0030 \pm 0.0026$&$ 0.0088\pm 0.0027$ & $0.0059\pm0.0013$ &$0.0127\pm0.0041$\\
 \hline\hline
 \end{tabular*}       
\end{table}

\begin{figure}
\centering
\includegraphics[width=\textwidth]{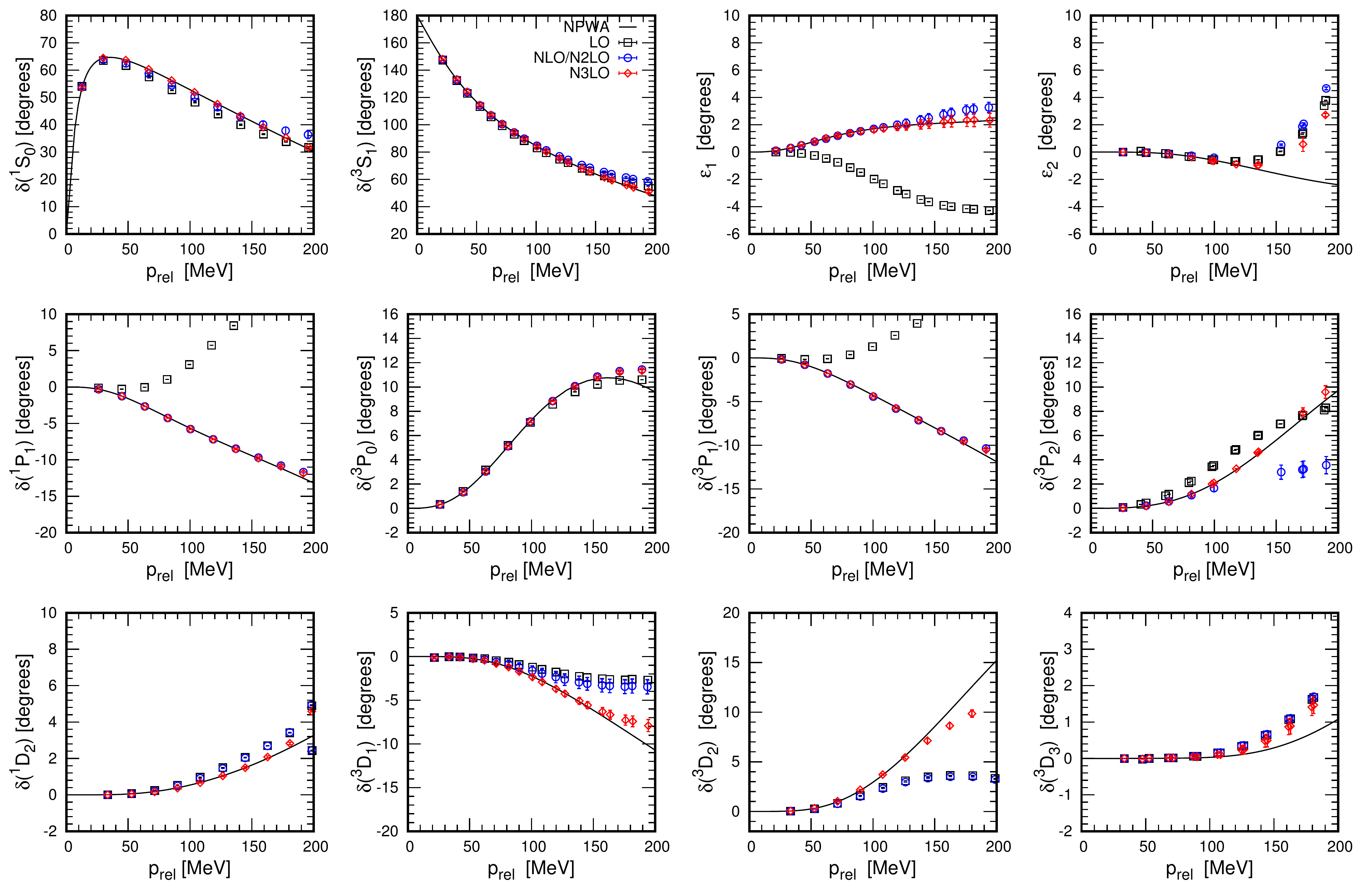}
\caption{(Color online) Neutron-proton scattering phase shifts and mixing angles versus relative momenta.   
The  lattice spacing $a = 1.97~{\rm fm}$ is used. The TPEP is not included explicitly as  discussed in the text.}\label{a100WO}
\end{figure}

In Fig.~(\ref{a100WO}) we show the phase shifts and mixing angles versus the relative momenta calculated using the  
coarsest lattice spacings, $a = 1.97~{\rm fm}$. We plot the results for relative momenta up to $p_{\rm rel} = 200~{\rm MeV}$. 
 The
error bars we quote in this plot and in the following plots indicate uncertainties
from the fitting procedure only.  A more comprehensive analysis that includes systematic
errors due to the truncated chiral EFT expansion is presented later in our
discussion. From the results, it is clear that with the new lattice operators the N$^3$LO calculations reproduce the NPWA phase shifts and mixing angles for most of 
the $S$, $P$, and $D$ waves with a good accuracy for relative momenta less than 200 MeV. One can also see clearly that the agreement improves with chiral order.
Unfortunately, the mixing angle $\epsilon_2$ bends up 
for the relative momenta $p_{\rm rel}$ at around 150 MeV, which indicates that higher order corrections, e.g., N4LO terms, 
or smaller lattice spacings would be needed to get the proper behavior for $\epsilon_2$ at higher momenta.     

\begin{figure}
\centering
\includegraphics[width=\textwidth]{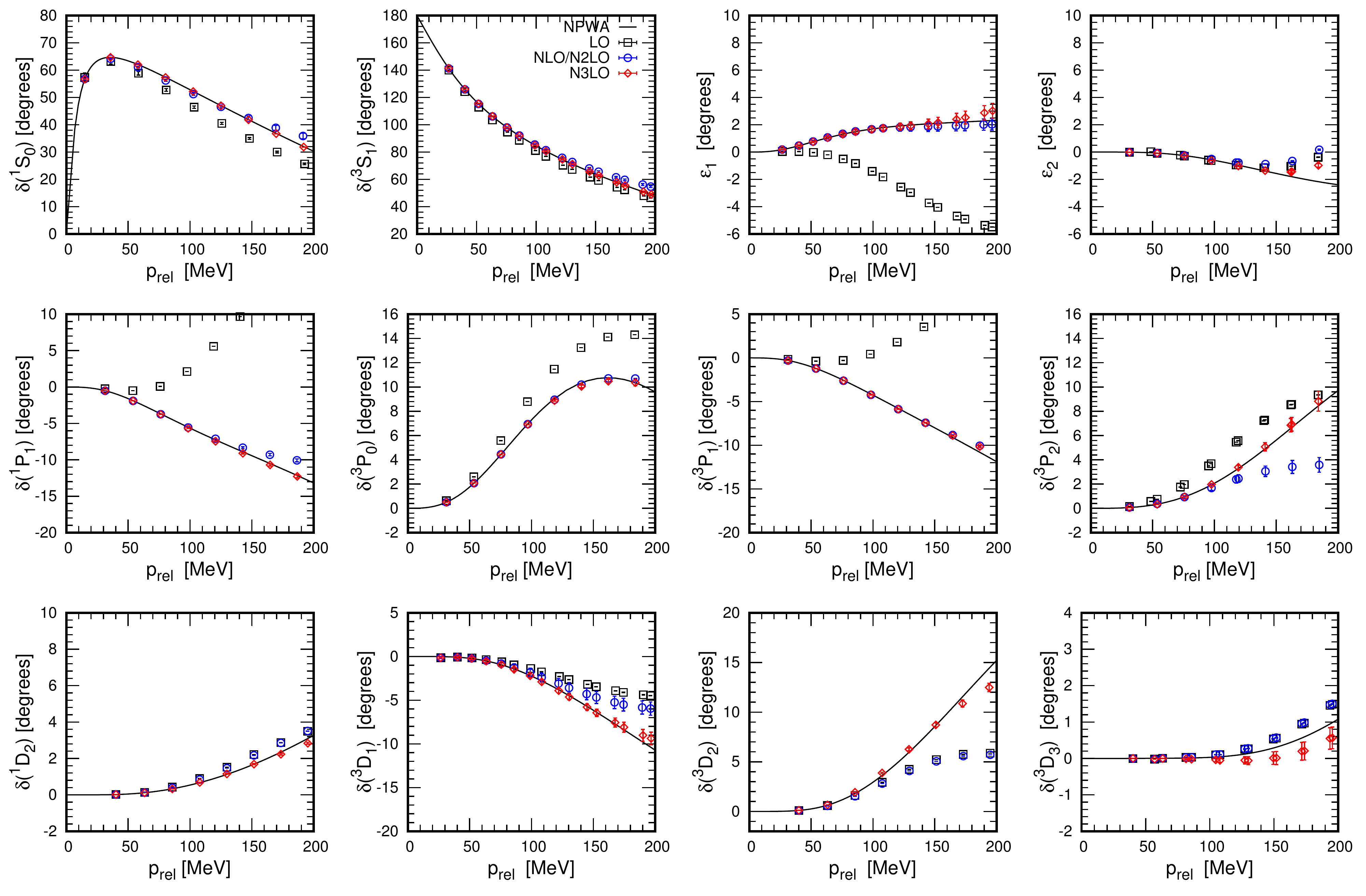}
\caption{(Color online) Neutron-proton scattering phase shifts and mixing angles versus relative momenta. 
The  lattice spacing $a = 1.64~{\rm fm}$ is used. The TPEP is not included explicitly
as  discussed in the text.}\label{a120WO}
\end{figure}

In Fig.~(\ref{a120WO}) we show the neutron-proton scattering phase shifts and mixing angles versus the relative momenta calculated using lattice spacing 
$a = 1.64~{\rm fm}$. The results are very similar to those at $a = 1.97~{\rm fm}$, but  the mixing angle $\epsilon_2$ stays accurate up to higher momenta compared 
with that at $a = 1.97~{\rm fm}$.  The smaller errors for the channels, $^3P_2$ and $\epsilon_2$, indicate
the results at $ a= 1.64~{\rm fm}$ have smaller lattice artifacts than those at $a = 1.97~{\rm fm}$, as one might expect. 

\begin{figure}
\centering
\includegraphics[width=\textwidth]{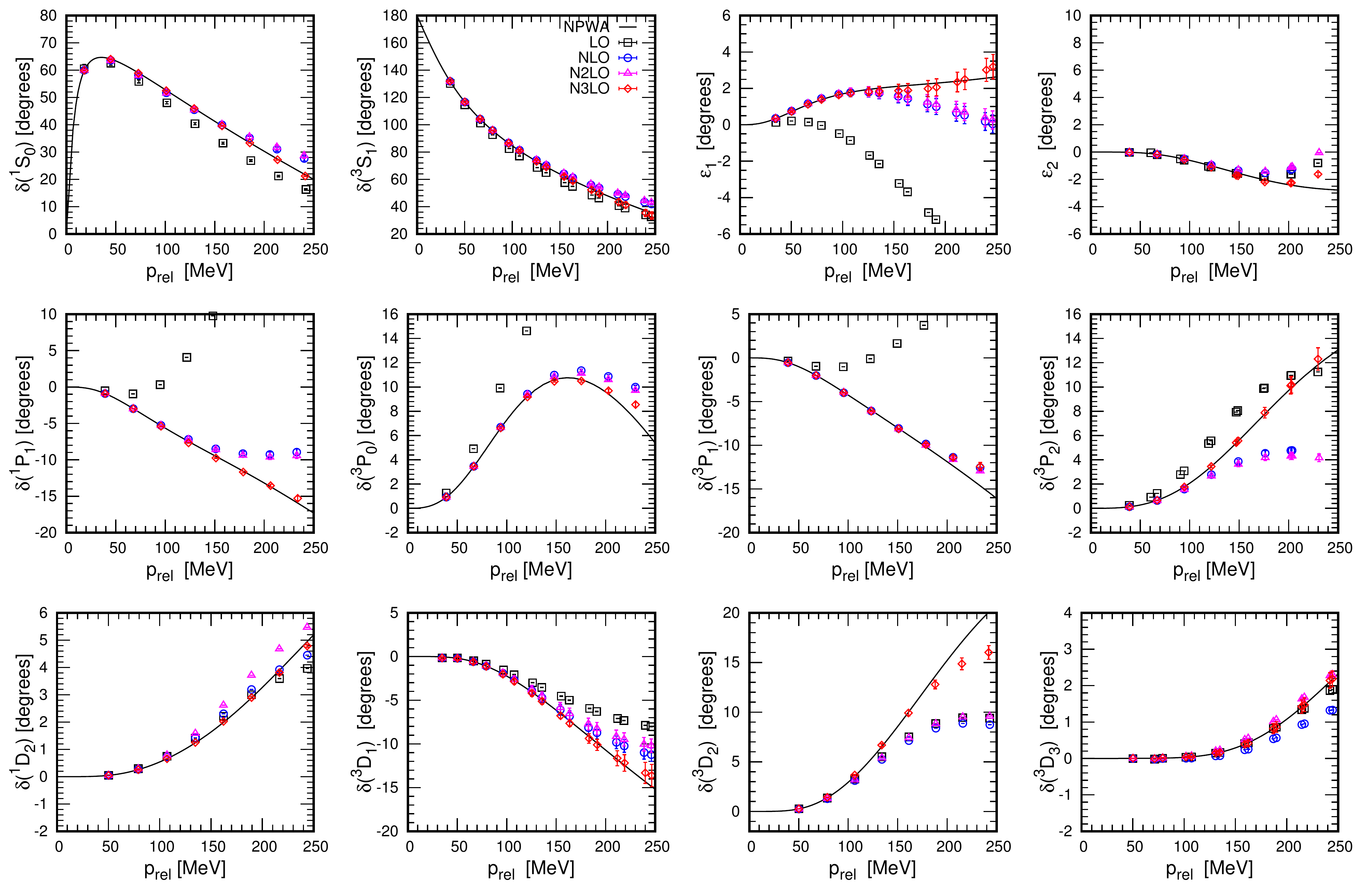}
\caption{(Color online) Neutron-proton scattering phase shifts and mixing angles versus relative momenta.  
The lattice spacing $a = 1.32~{\rm fm}$ and the full NN interactions are used. }\label{a150W}
\end{figure}

\begin{figure}
\centering
\includegraphics[width=\textwidth]{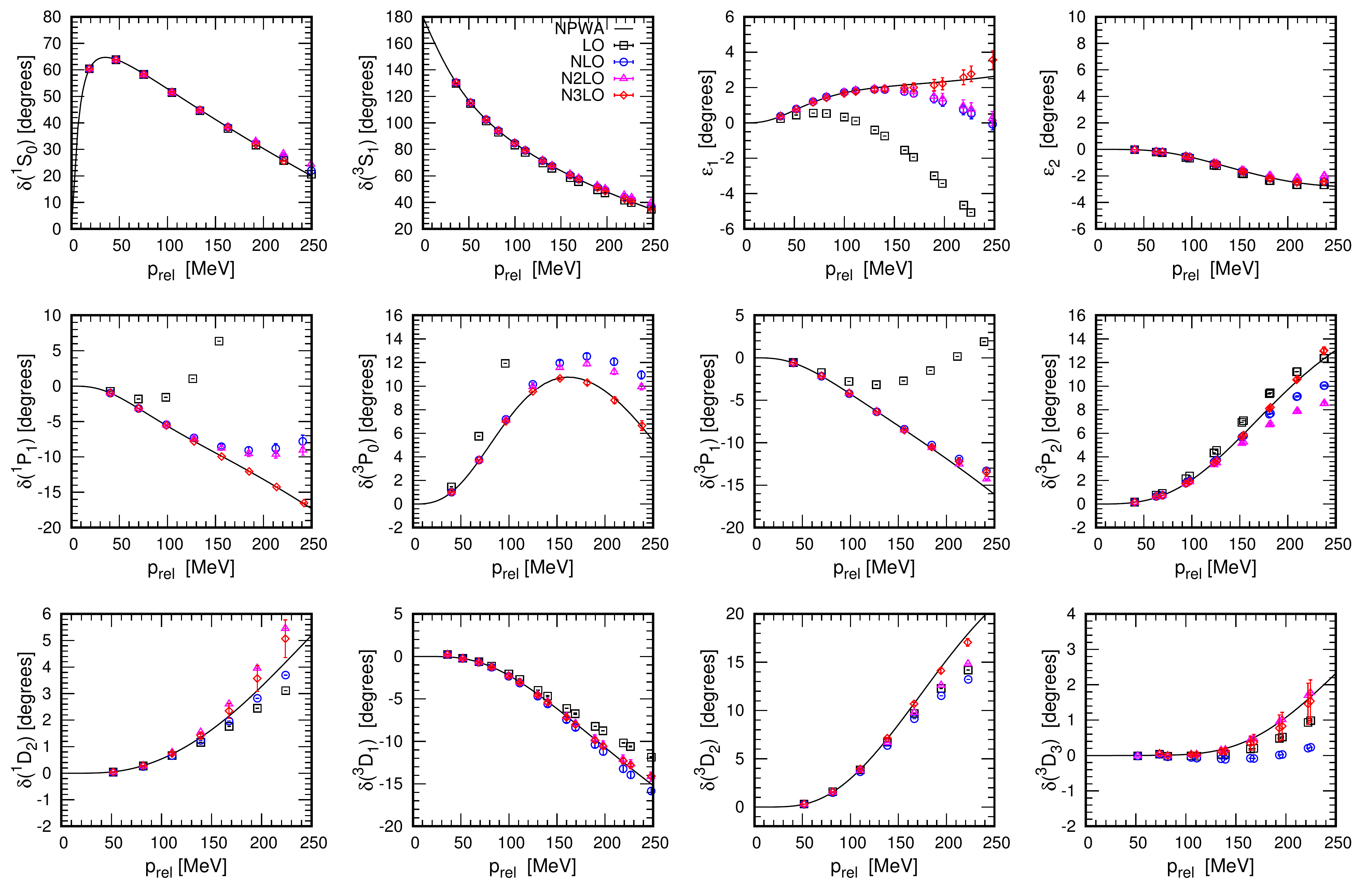}
\caption{(Color online) Neutron-proton scattering phase shifts and mixing angles  versus relative momenta.  
The lattice spacing $a = 0.99~{\rm fm}$ and the full NN interactions are used.}\label{a200W}
\end{figure}

For calculations involving the two smaller lattice spacings, $a = 1.32$ and $0.99~{\rm fm}$, we use the full NN interactions up to chiral order 
$O(Q^4)$ or N$^3$LO. The results are presented in Figs. (\ref{a150W}) and (\ref{a200W}), respectively.  We plot the results 
for relative momenta up to $p_{\rm rel} = 250~{\rm MeV}$.  Compared to the results using the larger lattice spacings, one can see 
clear improvement. Again good convergence is observed with increasing chiral order. With the full NN interactions up to order $O(Q^4)$, the calculation 
using $a = 0.99~{\rm fm}$ can describe the 
$S$, $P$, and $D$ waves with a good accuracy over the whole momentum range, $0 < p_{\rm rel} < 250~{\rm MeV}$. 

\begin{figure}
\centering
\includegraphics[width=\textwidth]{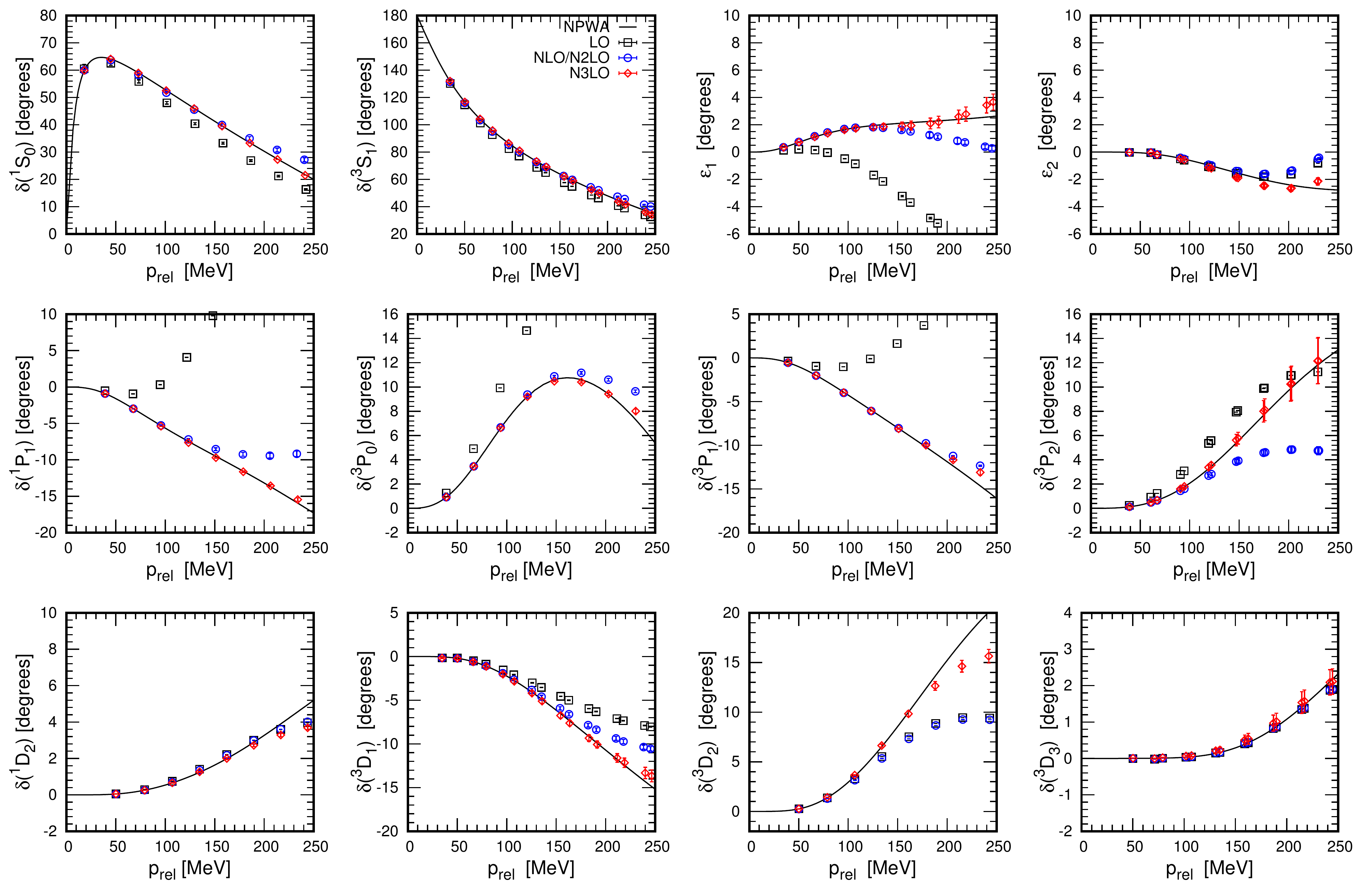}
\caption{(Color online) Neutron-proton scattering phase shifts and mixing angles  versus relative momenta.  
The lattice spacing $a = 1.32~{\rm fm}$ is used. The TPEP is not included in this case for comparison.}\label{a150WO}
\end{figure}

\begin{figure}
\centering
\includegraphics[width=\textwidth]{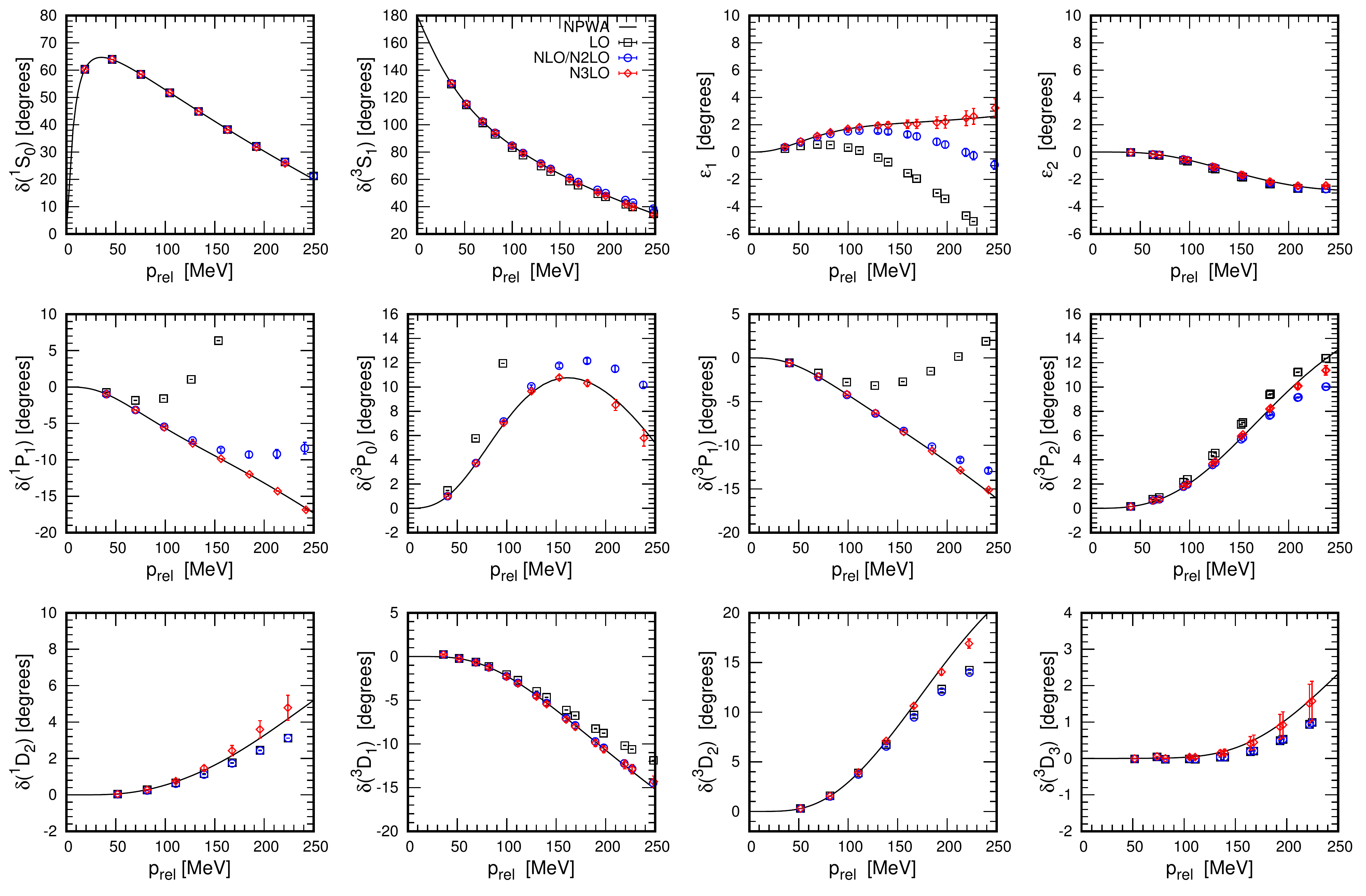}
\caption{(Color online) Neutron-proton scattering phase shifts and mixing angles  versus relative momenta. 
The lattice spacing $a = 0.99~{\rm fm}$ is used. The TPEP is not included in this
case for comparison.}\label{a200WO}
\end{figure}

To study the importance of the long-range part of the TPEP in the calculations, we also redo the same fits
without the TPEP for $a = 1.32$ and $a = 0.99~{\rm fm}$. Our results are shown in Fig. (\ref{a150WO}) and (\ref{a200WO}).
For the calculations using $a = 1.32$ and $a = 0.99~{\rm fm}$, the phase shifts and mixing angles without the TPEP 
are very similar to those with the TPEP, though the LECs are quite different. This indicates that the TPEP can be emulated 
by a retuning of the LECs. At the rather low scattering energies we probe, we do not see a clear improvement due to TPEP 
from the phase shifts and mixing angles.  However, we do expect that this will change at higher scattering energies.

\section{Properties of the deuteron}
\begin{figure}
\centering
\includegraphics[width=0.8\textwidth]{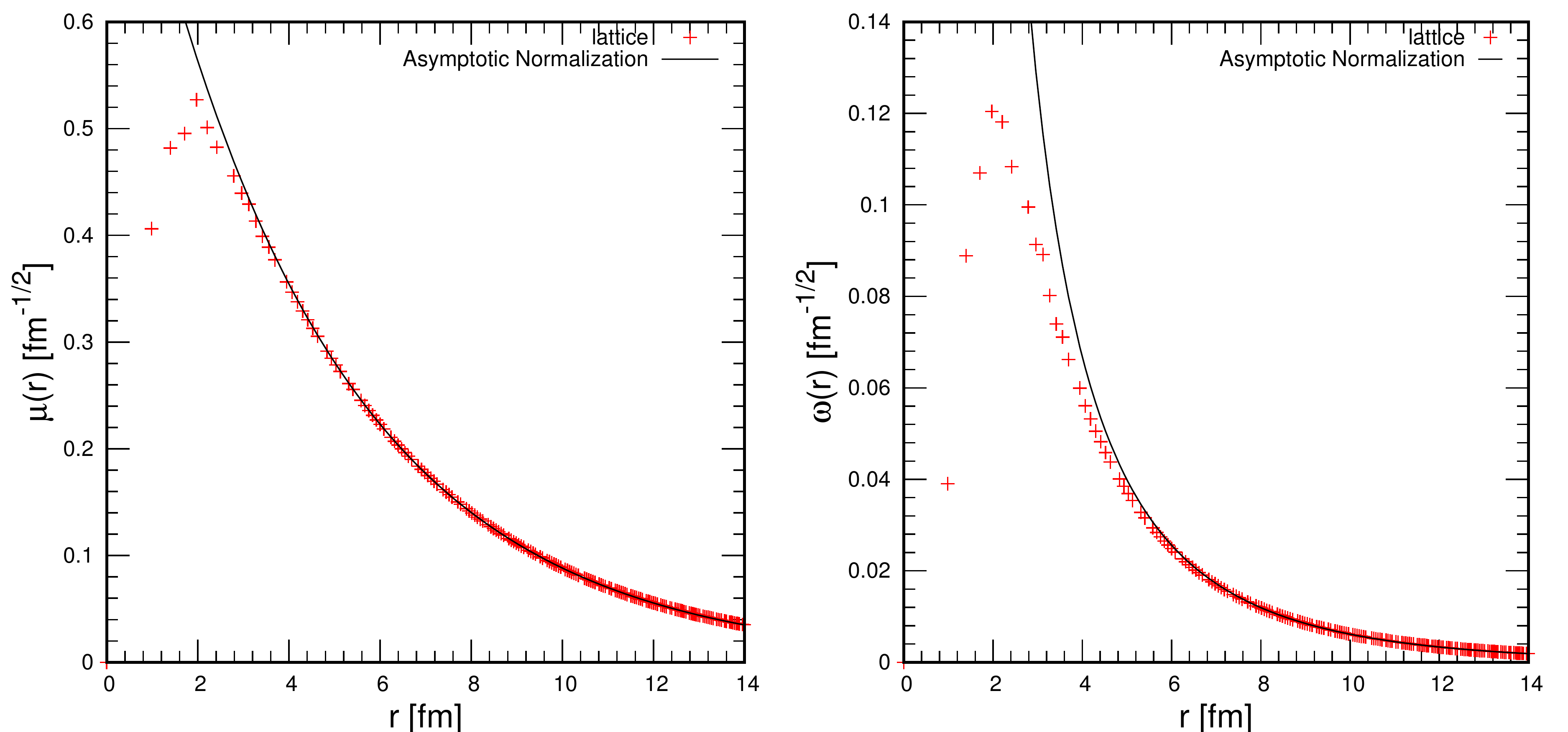}
\caption{(Color online) Radial wave functions of the deuteron and its asymptotic behavior at large $r$. Wave functions are 
calculated using $a =0.99~{\rm fm}$ with the full NN interaction up to order $O(Q^4)$. Left: $S$ wave. Right: $D$ wave. }
\label{funa200}
\end{figure}

In this section, we calculate the properties of the deuteron using the radial deuteron wave function obtained in the calculations with  $a = 0.99~{\rm fm}$ 
and the full NN interactions up to order $Q^4$ or N$^3$LO. At distance $r$ beyond the range of the interaction, the radial wave function for the deuteron in the ${}^3S_1$ channel behaves as,
\begin{eqnarray}
u(r) = A_S e^{-\gamma r}, \label{Swave}
\end{eqnarray} 
where $A_S$ is the $S$-wave asymptotic normalization coefficient.  Here $\gamma = \sqrt{m|E_d|}$ with $E_d$ denoting the deuteron binding energy. In the ${}^3D_1$ channel, the radial wave 
function behaves as 
\begin{eqnarray}
w(r) = \eta A_S \left[1 + \frac{3}{\gamma r} + \frac{3}{(\gamma r)^2} \right] e^{-\gamma r}. \label{Dwave}
\end{eqnarray} 
In Fig. (\ref{funa200}), we show the radial wave functions  of the deuteron calculated using $a = 0.99~{\rm fm}$ with the full NN 
interaction up to chiral order $O(Q^4)$. The left panel shows the $S$-wave radial wave function while the right panel is for the $D$-wave radial wave function.  In calculating 
the asymptotic normalization factors, we take the range  $8 < r < 14$ fm. From the plots, one can see clearly that when 
the neutron and proton are well separated the $S$ and $D$ waves behave as the asymptotic forms in Eq. (\ref{Swave}) and (\ref{Dwave}) respectively. 
The numerical values for $A_S$ and $\eta$ are shown in Table \ref{property:deuteron}.
\begin{table}
\renewcommand{\arraystretch}{1.3}
\centering
\caption{Deuteron properties and $S$-wave parameters calculated with the full NN interaction  up to chiral order $O(Q^4)$ using 
$a = 0.99$~fm. Error bars we list here indicate uncertainties from the fitting procedure only. }\label{property:deuteron}
\begin{tabular*}{\textwidth}{@{\extracolsep{\fill}}c|ccccr}
\hline\hline 
                                             & LO                           &  NLO                                    &  N$^2$LO                      & N$^3$LO                           & Empirical~~~ \\
\hline
$E_d$ (MeV)                        & $2.2246\pm0.0002$  &$2.224575\pm 0.000016$ &$2.224575\pm0.000025$& $2.224575\pm 0.000011$  & 2.224575(9)\cite{VanDerLeun:1982bhg} \\ 
$A_s(\mathrm{fm}^{-1/2})$   & $0.8662\pm0.0007$  &$0.8772\pm0.0003$          &$0.8777\pm0.0004$        &  $0.8785\pm0.0004$          & 0.8846(9)\cite{Ericson:1982ei} \\
$\eta$                                   & $0.0212\pm 0.0000$ &$0.0258\pm0.0001$          &$0.0257\pm0.0002$        &  $0.0254\pm0.0001$          &0.0256(4) \cite{Rodning:1990zz}\\
$Q_d(\mathrm{fm}^2)$         & $0.2134\pm0.00000$&$0.2641\pm0.0016$          &$0.2623\pm0.0023$        &  $0.2597\pm0.0013$           &0.2859(3) \cite{Bishop:1979zz} \\
$r_d$ (fm)                             &$1.9660\pm 0.0001$   &$1.9548\pm0.0005$          &$1.9555\pm0.0008$        &   $1.9545\pm0.0005$         & 1.97535(85) \cite{Huber:1998zz}\\
$a_{^3S_1}$                         &$ 5.461\pm0.000$      &$5.415\pm0.001$              &$5.421\pm0.002$            &   $5.417\pm 0.001$             & 5.424(4) \cite{Dumbrajs:1983jd}\\
$r_{^3S_1}$                         &$1.831\pm0.0003$     & $1.759\pm0.002$             &$1.760\pm0.003$            &    $1.758\pm0.002$              &  1.759(5)\cite{Dumbrajs:1983jd} \\
$a_{^1S_0}$                         & $-23.8\pm0.1$          & $-23.69\pm0.05$               &$-23.8\pm0.2$                & $-23.678\pm0.038$              & $-23.748(10)$\cite{Dumbrajs:1983jd} \\
$r_{^1S_0}$                          & $2.666\pm0.001$     & $2.647\pm0.003$             &$2.69\pm0.02$                 & $2.647\pm0.004 $                & $2.75(5)$ \cite{Dumbrajs:1983jd}\\
$P_D(\%)$                            & $1.92 $                      & $3.48  $                           & $3.41 $                            & $3.36  $                            &  \\
\hline\hline
\end{tabular*}
\end{table}

Using the radial wave functions, we can compute the root-mean-square radius of the deuteron, 
\begin{eqnarray}
r_d = \frac{1}{2}\left[\sum \delta r r^2 \left[u^2(r) + w^2(r)\right] \right]^{1/2}, 
\end{eqnarray}
where $\delta r $ is the small separation between the  radial bins we are using for
the radial deuteron wave function, and the summation is over all the radial
bins. In the same manner, we can also compute the  quadrupole moment of the deuteron, 
\begin{eqnarray}
Q_d = \frac{1}{20}\sum \delta r r^2 w(r) \left[\sqrt{8} u(r) - w(r) \right], 
\end{eqnarray}
 
In addition to the deuteron properties, we can also compute the $S$-wave effective range parameters at very low energies. 
The effective range expansion has the form

\begin{eqnarray}
p \cot (\delta) = -\frac{1}{a} + \frac{1}{2} r p^2 + O(p^4), 
\end{eqnarray}
where $p$ is the relative momenta between the neutron and the proton, while $a$
and $r$ are the scattering length and effective range respectively. 
 Using these formula, we can extract the scattering length and effective range
for $^3S_1$ and $^1S_0$. 

In Table \ref{property:deuteron}, we present the properties of the deuteron and $S$-wave parameters obtained using the 
$a = 0.99~{\rm fm}$ and the  full NN interactions up to order $Q^4$. In order to provide some
insight into the nature of the lattice wave functions, we also list the computed $D$-wave probabilities of the deuteron, $P_D$.  We note, however, that $P_D$ is strongly dependent on short-distance physics and the scale at which it is regulated \cite{Friar:1979zz}.

From the results in Table~\ref{property:deuteron}, it is clear that the deuteron
properties can be reproduced accurately at lattice spacing $ a = 0.99~{\rm
fm}$. There are, however, still some small 
systematic discrepancies that suggest additional corrections are needed.
 While these could be due to corrections beyond N$^3$LO in the lattice Hamiltonian,
they could also be due to missing corrections to the observables themselves
such as the $r^2$ operator.  Such corrections are needed to 
cancel ambiguities on how the operators are defined on a discrete lattice.
 For example, the nucleons could be regarded as exactly localized as delta
functions at the lattice sites or they could be viewed as having some other
distribution with a width comparable to the lattice spacing.  While numerically
small, these corrections
to the operator observables are required for a full accounting of all lattice
and regularization artifacts.
 See, for example, Ref.~\cite{Klein:2018iqa}.  This is an interesting but extensive subject
that requires further
investigation 
in future studies.

\section{Theoretical uncertainties}
It is necessary also to address the convergence of the effective field
theory expansion on the lattice and their associated systematic errors.  These important topics have generated much recent interest 
\cite{Epelbaum:2014efa,Epelbaum:2014sza,Furnstahl:2014xsa,Furnstahl:2015rha}. We follow the prescription in Refs.~\cite{Epelbaum:2014efa,Epelbaum:2014sza} where the theoretical uncertainty for some observable $X(p)$ at order N$^m$LO and momentum $p$ is given by
\begin{equation}
\Delta X^{{\rm N}^m{\rm LO}}(p) = \max\left( Q^{m+2}\left| X^{\rm LO}(p) \right|,
Q^{m}\left| X^{\rm LO}(p) - X^{\rm NLO}(p)  \right|, 
\cdots, Q^1 \left| X^{{\rm N}^{m-1}{\rm LO}}(p) - X^{{\rm N}^m{\rm LO}}(p)  \right| \right) . 
\end{equation} 
Here $Q$ is the estimated expansion parameter controlling the rate of convergence,
\begin{equation}
Q = \max \left( p/\Lambda_b,M_{\pi}/{\Lambda_b} \right),
\end{equation}
and $\Lambda_b$ the breakdown momentum scale. On the lattice, cubic symmetry replaces the rotational symmetry of
the continuum, and the $(2L + 1)$-dimensional irreducible representation of
SO(3) decomposes into irreducible representations 
of the rotational octahedral group O. For example, $L=0$ corresponds to the
$A_1$ of O, and  $L = 1$ corresponds to the $T_1$ of O. However 
$L=2$ splits into the $E$ and $T_2$ representations of O, and similar splittings
occur in all of the larger $L$ representations.  As a result the breaking
of rotational symmetry for $L \ge 2$ is numerically larger than 
that for $L<2$. This leads to a lower momentum breakdown scale for $D$ waves
and above compared to the $S$ and $P$ waves. To account for this in our 
calculations, we take $\Lambda_b$ 
to be the lattice momentum cutoff $\Lambda_{\rm latt} = \pi/a$ for the lower partial waves, and we take $\Lambda_b = (2/3)\Lambda_{\rm latt}$ for $\epsilon_2$,
$D$ waves, and higher partial waves.

\begin{figure}
\centering
\includegraphics[width=\textwidth]{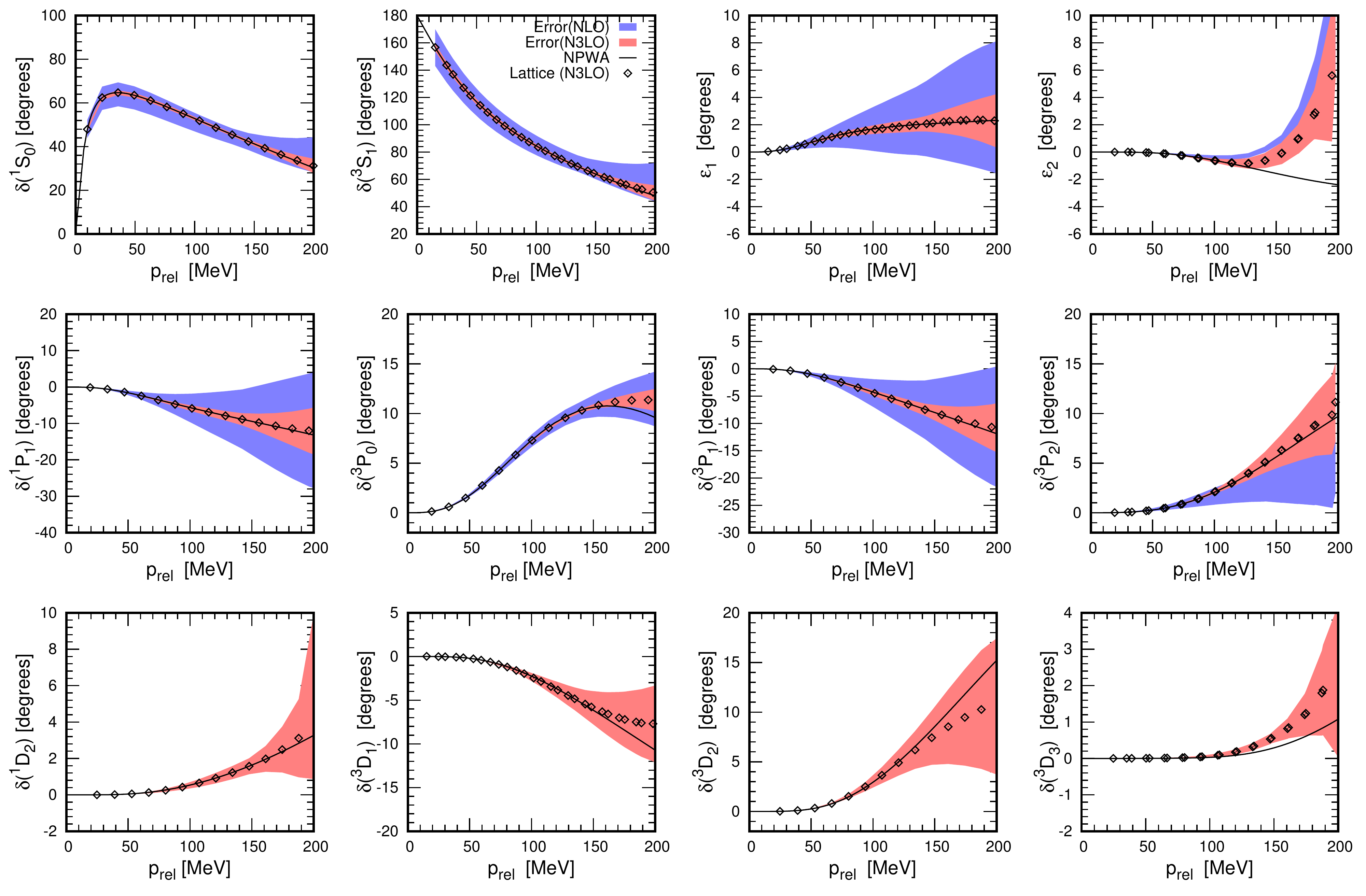}
\caption{(Color online) Theoretical error bands for the neutron-proton scattering phase shifts and mixing angles versus the relative momenta 
for $a = 1.97~{\rm fm}$. Blue and red bands signify the estimated uncertainties at NLO and N$^3$LO respectively. The black solid line and diamonds 
denote the phase shift or mixing angle from the Nijmegen partial-wave analysis (NPWA) and lattice calculation at N$^3$LO, respectively.}
\label{ErrorWOa100}
\end{figure}

\begin{figure}
\centering
\includegraphics[width=\textwidth]{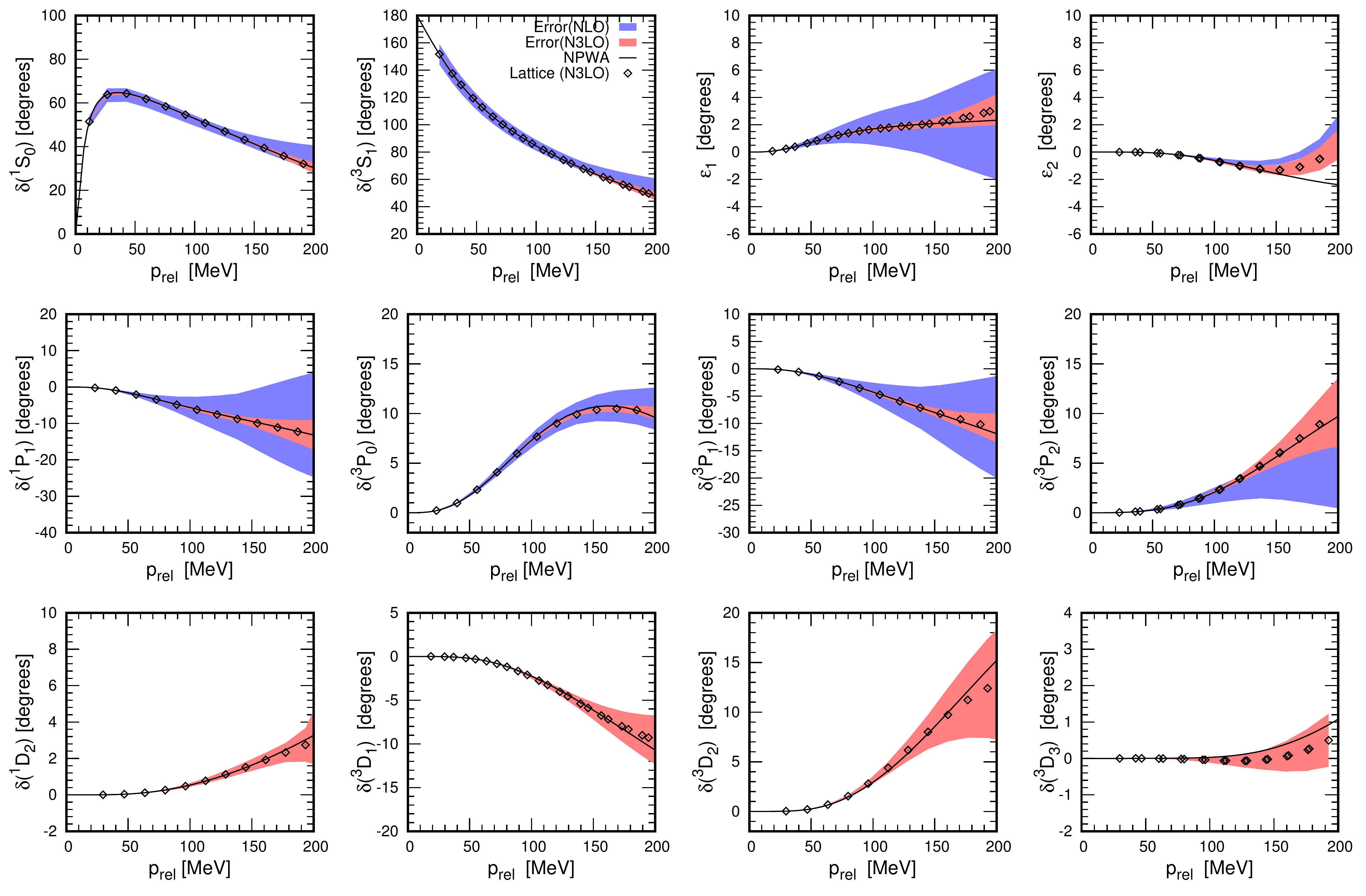}
\caption{(Color online) Theoretical error bands for the neutron-proton scattering phase shifts and mixing angles versus the relative momenta 
for $a = 1.64~{\rm fm}$. Blue and red bands signify the estimated uncertainties at NLO and N$^3$LO, respectively. The black solid line and diamonds 
denote phase shift or mixing angle from the Nijmegen partial-wave analysis (NPWA) and lattice calculation at N$^3$LO, respectively. }
\label{ErrorWOa120}
\end{figure}

\begin{figure}
\centering
\includegraphics[width=\textwidth]{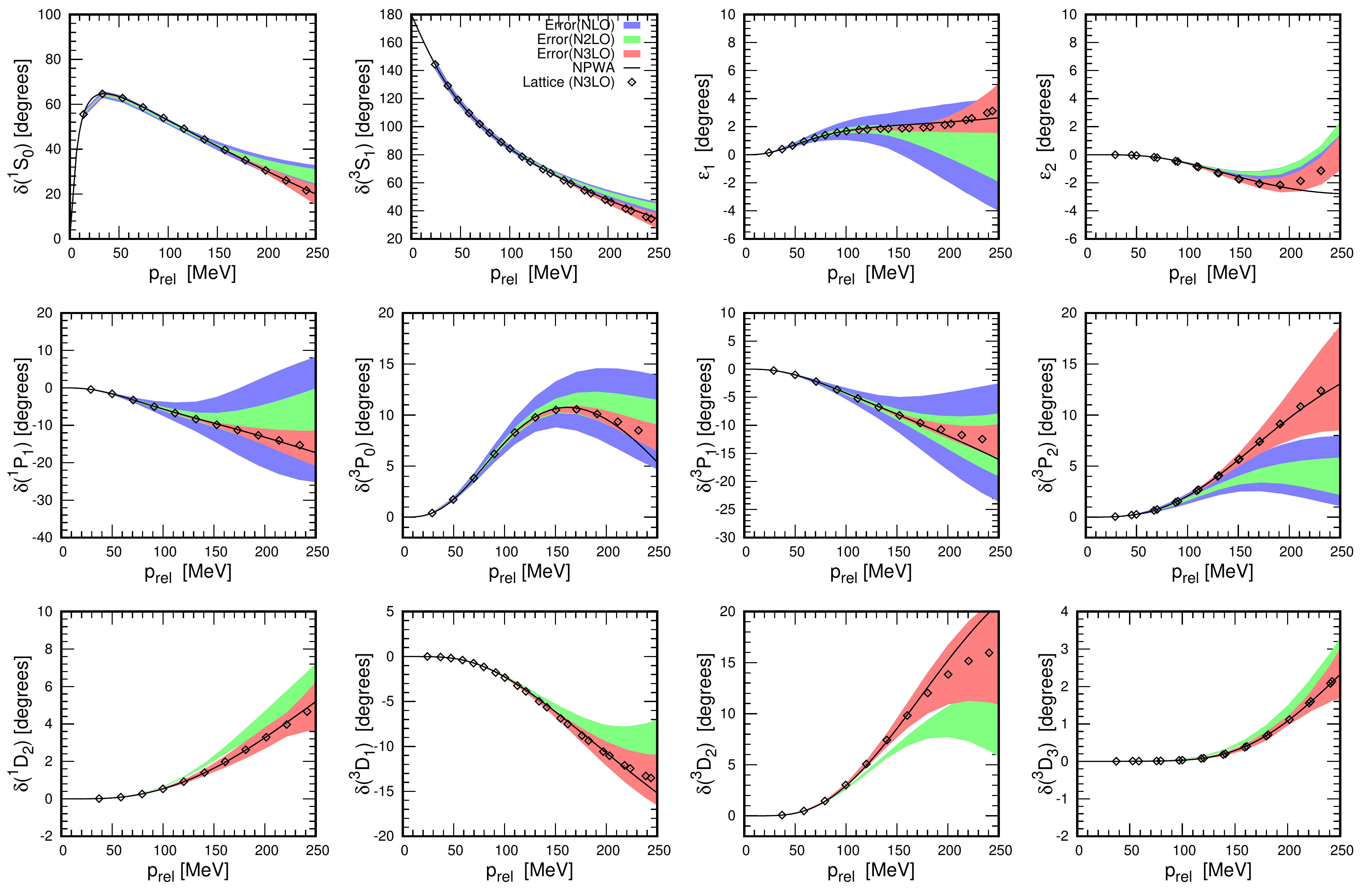}
\caption{(Color online) Theoretical error bands for neutron-proton scattering phase shifts and mixing angles versus the relative momenta 
for $a = 1.32~{\rm fm}$ with the full NN interaction. 
Blue, green, and red bands signify the estimated uncertainties at NLO, N$^2$LO, and N$^3$LO, respectively. The black solid line and diamonds 
denote the phase shift or mixing angle from the Nijmegen partial-wave analysis (NPWA) and lattice calculation at N$^3$LO, respectively. }
\label{Errora150}
\end{figure}

\begin{figure}
\centering
\includegraphics[width=\textwidth]{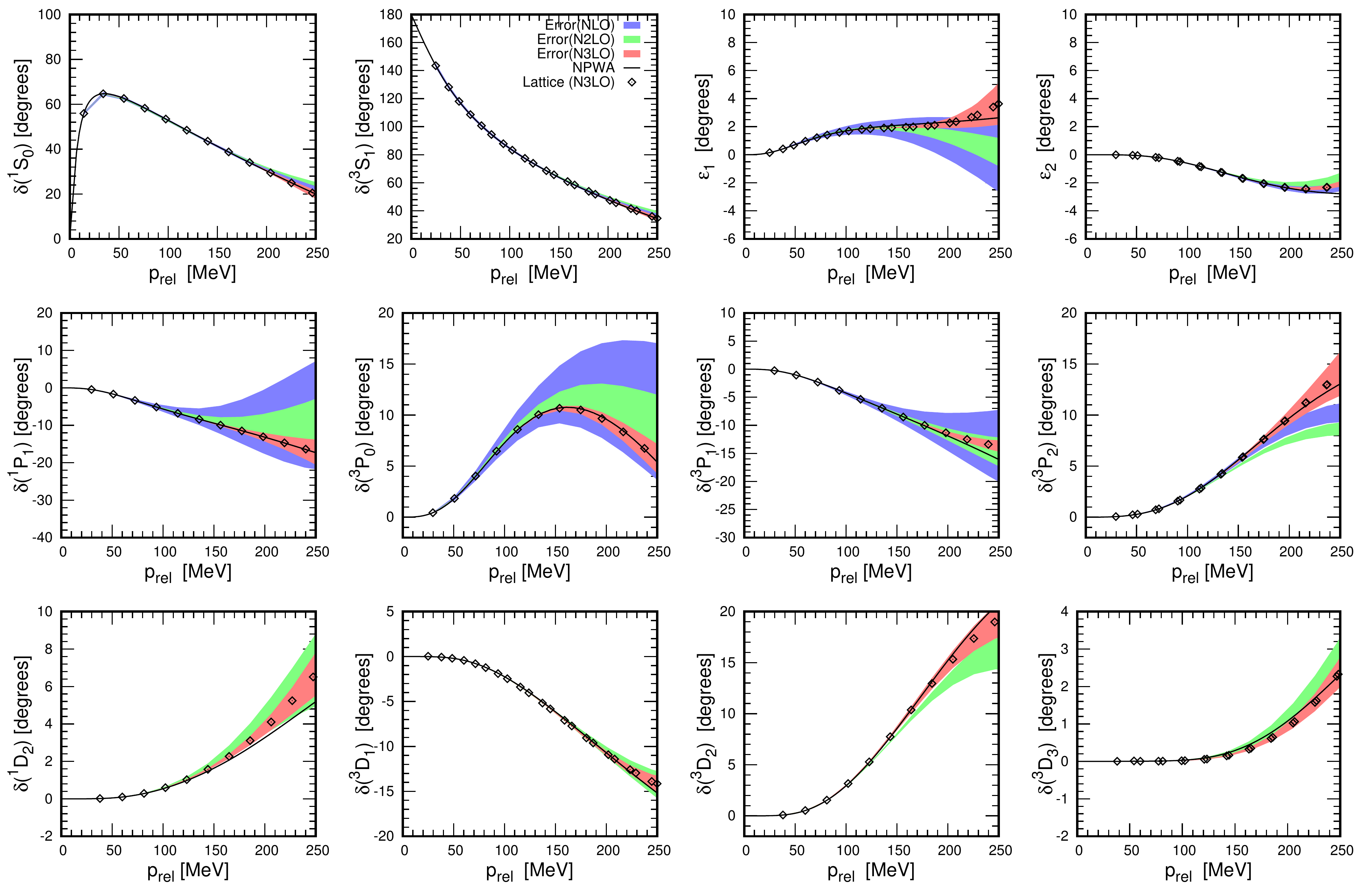}
\caption{(Color online) Theoretical error bands for the neutron-proton scattering phase shifts and mixing angles versus the relative momenta 
for $a = 0.99~{\rm fm}$ with the full NN interaction. 
Blue, green, and red bands signify the estimated uncertainties at NLO, N$^2$LO and N$^3$LO respectively. The black solid line and diamonds 
denote the phase shift or mixing angle from the Nijmegen partial-wave analysis (NPWA) and lattice calculation at N$^3$LO, respectively. }
\label{Errora200}
\end{figure}

\begin{figure}
\centering
\includegraphics[width=\textwidth]{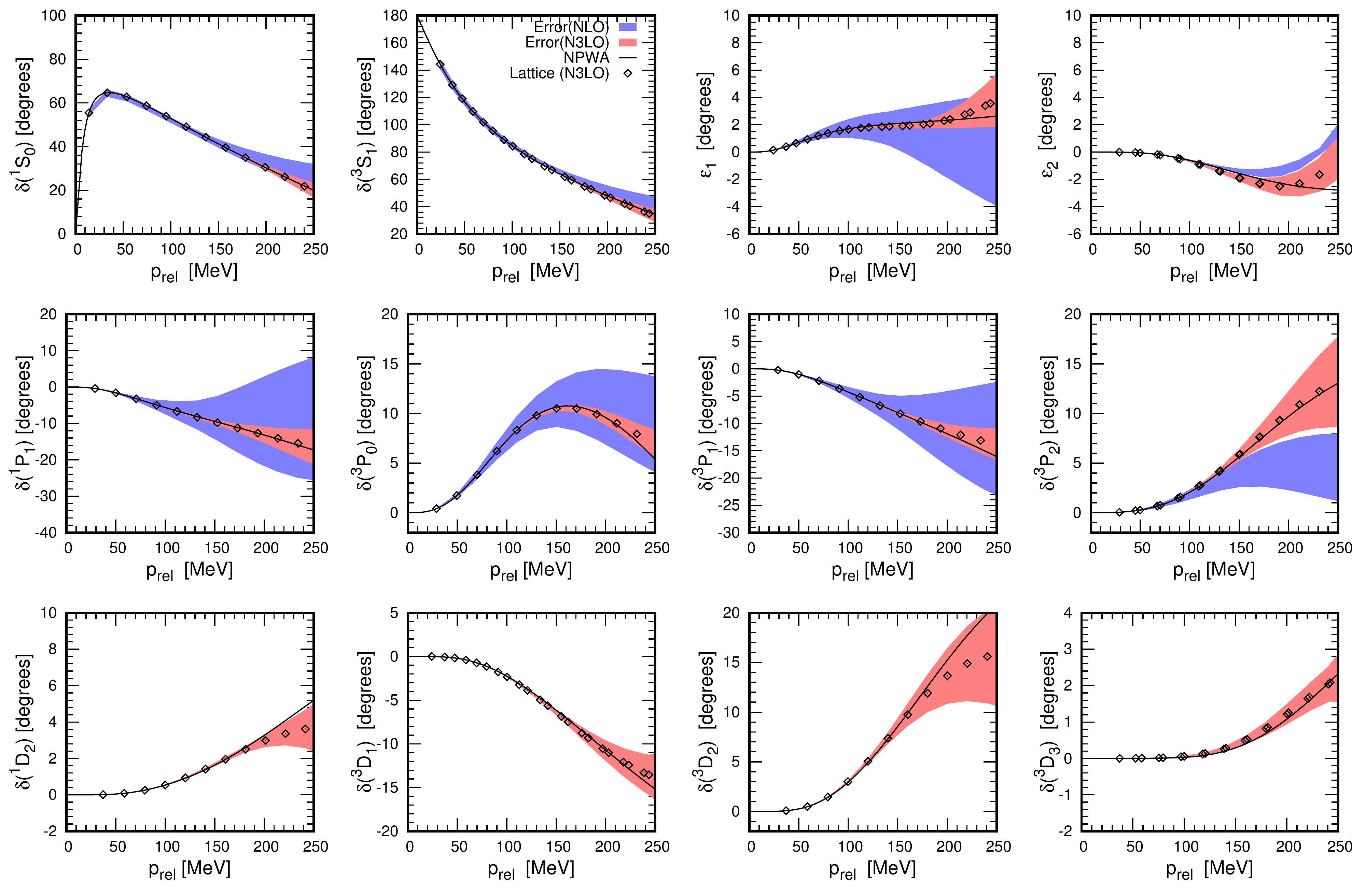}
\caption{(Color online) Theoretical error bands for the neutron-proton scattering phase shifts and mixing angles versus the relative momenta 
for $a = 1.32~{\rm fm}$ without the TPEP. 
Blue and red bands signify the estimated uncertainties at NLO and N$^3$LO respectively. The black solid line and diamonds 
denote phase shift or mixing angle from the Nijmegen partial-wave analysis (NPWA) and lattice calculation at N$^3$LO, respectively. }
\label{ErrorWOa150}
\end{figure}

\begin{figure}
\centering
\includegraphics[width=\textwidth]{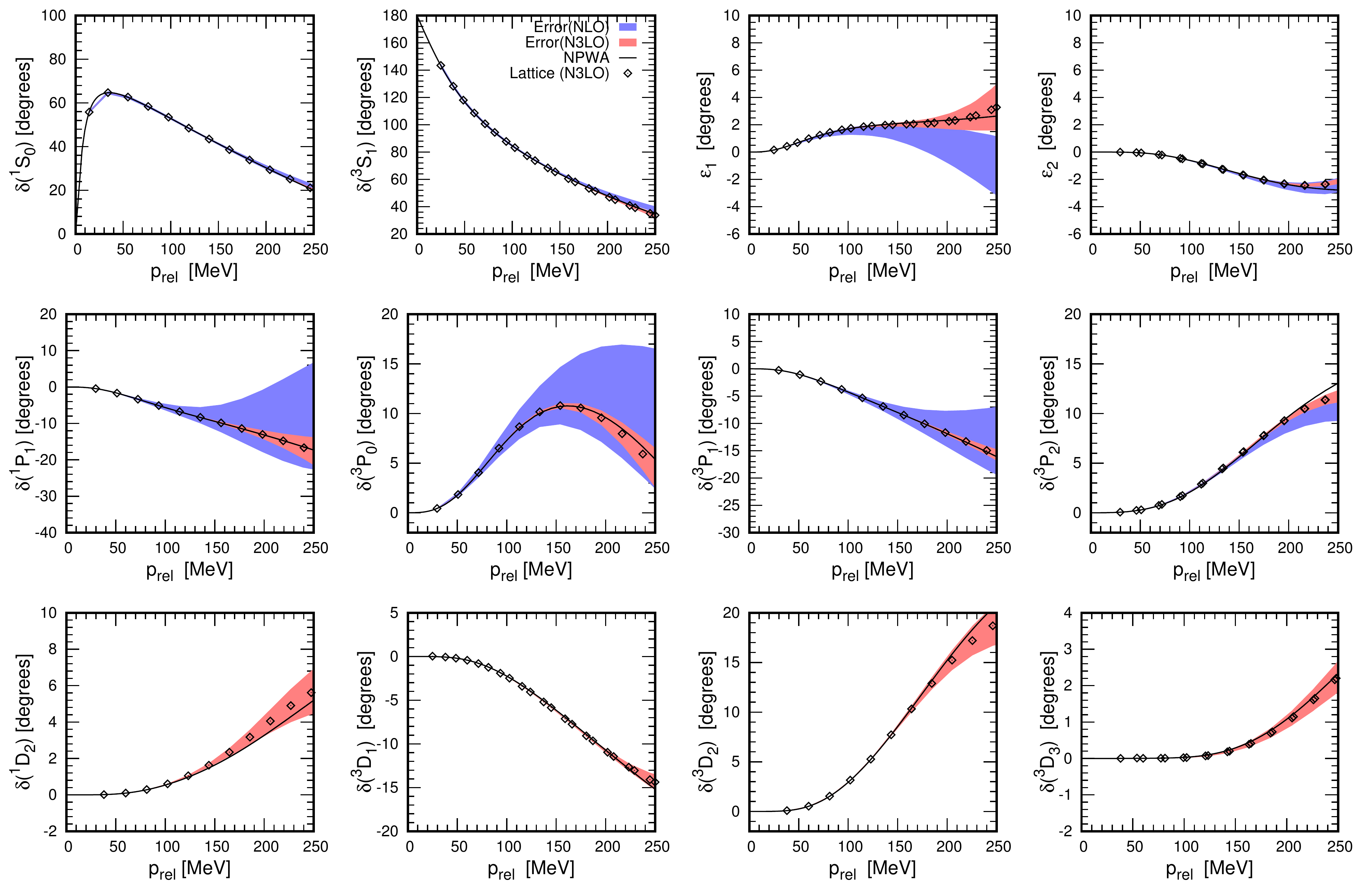}
\caption{(Color online) Theoretical error bands for the neutron-proton scattering phase shifts and mixing angles versus the relative momenta 
for $a = 0.99~{\rm fm}$ without the TPEP. 
Blue and red bands signify the estimated uncertainties at NLO and N$^3$LO respectively. The black solid line and diamonds 
denote phase shift or mixing angle from the Nijmegen partial-wave analysis (NPWA) and lattice calculation at N$^3$LO, respectively. }
\label{ErrorWOa200}
\end{figure}

We will study the dependence of the lattice breakdown scale on $L$ and $J$  in more 
detail in future work. The theoretical error bands for the neutron-proton scattering phase shifts and mixing
angles versus the relative momenta for $a = 1.97$, $1.64$, $1.32$ and $0.99~{\rm fm}$ are shown in Figs.~(\ref{ErrorWOa100}-\ref{Errora200}), 
in which we see a systematic decrease in the uncertainties for the $S$- and $D$- wave phase shifts with decreasing lattice 
spacing. The unexpectedly small NLO uncertainties for the ${}^3P_0$ phase shifts at a coarse lattice spacing are caused by 
the rather good but accidental accuracy of the  ${}^3P_0$ phase shifts at LO.  
We also show the estimated theoretical uncertainties for the neutron-proton scattering 
phase shifts and mixing angles for $a = 1.32$ and $0.99~{\rm fm}$ without the long-range TPEP in Figs.~(\ref{ErrorWOa150}) 
and (\ref{ErrorWOa200}). With only a few exceptions, the error bands for each order generally overlap with each other and 
cover the empirical phase shifts.  This is a promising sign of convergence of the chiral effective field theory expansion on the lattice.

\section{Summary and outlook}\label{summary}

We have proposed a new lattice formulation of the chiral NN force which is easily decomposed into partial waves. The new 
lattice operators work as projection operators, which only survive in particular channels. This advantage greatly simplifies the 
fitting procedure. Instead of fitting the phase shifts and mixing angles for all the channels simultaneously, 
only one uncoupled channel or two coupled channels are needed to be computed for each calculation.  

To study the dependence on the lattice spacing, we have computed neutron-proton phase shifts and mixing angles 
using  four different lattice spacings, $a = 1.97~{\rm fm}$, $1.64~{\rm fm}$, $1.32~{\rm fm}$ 
and $0.99~{\rm fm}$. For two  coarser lattice spacings, $a = 1.97~{\rm fm}$ or $1.64~{\rm fm}$, we did not explicitly include the TPEP, 
whereas for those using $a = 1.32~{\rm fm}$ or $0.99~{\rm fm}$, we did. 
Our numerical results indicate  a good convergence with respect to chiral order. One also observes an obvious
improvement when the lattice spacing is decreased  Comparing the results obtained with and without the TPEP at 
lattice spacings of $ a = 1.32$ and $0.99~{\rm fm}$, we did not find significant differences.
This may, however, be a consequence of the rather low scattering energies we probe,
and  we expect that differences would appear at higher scattering energies. 

We have also studied the properties of the deuteron wave function and the  $S$-wave effective
range parameters obtained 
with the full NN interaction at lattice spacing $a = 0.99~{\rm fm}$. The numerical values are very close to the 
empirical values, which indicates that the current version of NN interactions is quite accurate, and a very significant 
improvement over previous lattice studies. Some small discrepancies remain, but these may well be fixed in studies 
that reach a higher order in the chiral effective field theory expansion.  
 
In summary, the new lattice interactions are far more efficient and accurate in reproducing physical data than 
previous lattice interactions.  We have begun studying the properties of light and medium-mass nuclei using these 
interactions, and the results are promising.  These interactions were specifically designed to facilitate very efficient  
Monte Carlo simulations of few- and many-body systems using auxiliary fields.  The results of these studies using 
these new interactions will be reported in several future publications.

\section{Acknowledgement }

We acknowledge partial financial support from the Deutsche Forschungsgemeinschaft (SFB/TR 110,
``Symmetries and the Emergence of Structure in QCD"), 
the BMBF (Grant No. 05P2015),
the U.S. Department of Energy (DE-SC0018638), and the Scientific and Technological Research 
Council of Turkey (TUBITAK project no. 116F400). Further support was provided by the Chinese 
Academy of Sciences (CAS) President's International Fellowship Initiative (PIFI) (Grant No. 2018DM0034) 
and by VolkswagenStiftung (Grant No. 93562).  The computational resources were provided by the J\"{u}lich 
Supercomputing Centre at Forschungszentrum J\"{u}lich, Oak Ridge Leadership Computing Facility, 
RWTH Aachen, North Carolina State University, and Michigan State University.

\bibliography{two_nucleon}{}
\bibliographystyle{apsrev}
  
\end{document}